\documentclass[journal=jctcce,manuscript=article]{achemso}
\setkeys{acs}{usetitle = true}
\setkeys{acs}{maxauthors=10,etalmode=truncate} \usepackage{xspace,
  amsmath, multirow, setspace, graphicx, subfigure} \usepackage{bm,
  dcolumn, epsfig, color, xr, booktabs, longtable, lscape}
\usepackage{subfig} \usepackage{tikz} \usepackage{mathtools}
\newcommand{\implyarrow}{%
\mathrel{\raisebox{1.5ex}{\rotatebox[origin=c]{90}{\mathhexbox37F}}}}

\usepackage{xr}

\title[]{Development of an Advanced Force Field for Water using Variational Energy Decomposition Analysis}

\author{Akshaya K. Das}
\author{Lars Urban}
\author{Itai Leven}
\author{Matthias Loipersberger}
\author{Abdulrahman Aldossary}
\author{Martin Head-Gordon}
\affiliation{ 
Pitzer Center for Theoretical Chemistry, Department of Chemistry, University of California, Berkeley CA 94720}
\author{Teresa Head-Gordon}
 \affiliation{
 Pitzer Center for Theoretical Chemistry, Departments of Chemistry, Bioengineering, Chemical and Biomolecular Engineering, University of California, Berkeley CA 94720
}
\email{thg@berkeley.edu}


\begin{document}

\begin{abstract}
Given the piecewise approach to modeling intermolecular interactions for force fields, they can be difficult to parameterize since they are fit to data like total energies that only indirectly connect to their separable functional forms. Furthermore, by neglecting certain types of molecular interactions such as charge penetration and charge transfer, most classical force fields must rely on, but do not always demonstrate, how cancellation of errors occurs among the remaining molecular interactions accounted for such as exchange repulsion, electrostatics, and polarization. In this work we present the first generation of the (many-body) MB-UCB force field that explicitly accounts for the decomposed molecular interactions commensurate with a variational energy decomposition analysis, including charge transfer, with force field design choices that reduce the computational expense of the MB-UCB potential while remaining accurate. We optimize parameters using only single water molecule and water cluster data up through pentamers, with no fitting to condensed phase data, and we demonstrate that high accuracy is maintained when the force field is subsequently validated against conformational energies of larger water cluster data sets, radial distribution functions of the liquid phase, and the temperature dependence of thermodynamic and transport water properties. We conclude that MB-UCB is comparable in performance to MB-Pol, but is less expensive and more transferable by eliminating the need to represent short-ranged interactions through large parameter fits to high order polynomials.\end{abstract}

\section{Introduction}
The field of molecular modeling has historically relied on a simple representation of the potential energy surface of molecules based on what is known as the "pairwise additive" approximation.\cite{Lifson1968} However, pairwise additive forces are limited when there is an "asymmetric environment"\cite{Remsing2014}, such as the heterogeneity of an interface\cite{Jungwirth2008,Vazdar2012}, anisotropic electric fields in native and synthetic enzyme active sites\cite{Vaissier2018a,Vaissier2018b}, the complex structural organization of  functional materials\cite{Vaissier2018c,McDaniel2012}, and the design of new drug molecules for which highly directional molecular interactions must be finely tuned for binding specificity\cite{XueweiLiu2018,DeVivo2016}. Undoubtedly, creating a higher accuracy force field for this range of system complexity requires the introduction of new functional forms that describe many-body interactions such as polarization and non-classical effects like charge penetration\cite{Demerdash2018,Rackers2017, Wang2015, Wang2014,Slipchenko2009, Wang2010, Stone2011, Piquemal2003, Freitag2000,Ngo2019,Wang2017jpcb,Wang2014jpcl}. Powerful success stories for improved accuracy in prediction are evident from advanced force fields that incorporate such physics, i.e. SIBFA (Sum of Interactions Between Fragments Ab initio computed)\cite{Piquemal2007b}, the CHARMM Drude model\cite{Lopes2013}, Atomic Multipole Optimized Energetics for Biomolecular Applications (AMOEBA)\cite{Ren2002,Pengyu2013,Ponder2010,Rackers2018}, Gaussian Electrostatic Model (GEM)\cite{Piquemal2006,Cisneros2006}, Effective Fragment Potential (EFP)\cite{Gordon2012,Slipchenko2009}, and MB-Pol\cite{Babin2012,Babin2013}. Even so, the greater complexity of this additional physics also poses greater challenges for rational force field design, as well as algorithmic and  software challenges needed to increase computational efficiency in order to be used effectively for sampling in the condensed phase.\cite{Demerdash2014,Albaugh2016} 

The rational design challenge is manifested in the fact that the advanced force fields are often not as transferable as desired, proving surprisingly brittle in blind-prediction challenges\cite{Mobley2012,Mobley2014,Bradshaw2016}, while other MB potentials such as MB-Pol disregard transferability in favor of explicitly parameterizing each new chemical system to ensure high accuracy with commensurate increases in computational expense.\cite{Babin2012,Babin2013} Another way forward to impact the force field design and sampling challenge is to systematically develop computationally tractable force field terms that match the decomposition of first principles quantum mechanical (QM) interaction energies. These piecewise contributions, including permanent and induced electrostatics, Pauli repulsion, dispersion, charge penetration, and in some cases charge transfer, can be revealed by chemically motivated energy decomposition analysis (EDA) methods.\cite{Kitaura1976,Chen1996,Mitoraj2009,Su2009,Reed1988,Jeziorski1994,Misquitta2005,Gao2012} There are already successful efforts in this direction such as the effective fragment potential (EFP) method\cite{Gordon2013,Gordon2009,Gordon2001}, the variational many-body expansion to guide the development of dispersion, exchange repulsion, and charge transfer contributions between different fragments in the X-Pol model of Gao and co-workers\cite{Gao2012}, force fields based upon symmetry-adapted perturbation theory (SAPT)\cite{Tafipolsky2016,McDaniel2012,Schmidt2015,Rackers2018,amoebaplus}, or that are analyzed through variational absolutely-localized-molecular-orbitals (ALMO)-EDA methods\cite{Mao2016,Demerdash2017,Das2018}. It is well-known that various EDA schemes all agree in the asymptotic region for most intermolecular interactions, but due to inherent non-separability of the QM energies into individual pieces, EDA schemes are non-unique at or near equilibrium and into the compressed region when electron density of the molecular fragments overlap.\cite{Phipps2015} Our view is that a force field that reproduces the piecewise energy decomposition of any particular EDA scheme in the form of a truncated many-body expansion, and in turn shows that it is transferable to describe a larger or more complex system or phase, is an important way to fully validate any particular EDA approach. 

In our previous work\cite{Mao2016,Demerdash2017,Das2018}, we compared piecewise decomposition of the AMOEBA water force field\cite{Ren2003} with ALMO-EDA at the level of $\omega$B97X-V/def2-QZVPPD\cite{Mardirossian2014,Rappoport2010} for various water and water-ions systems \cite{Demerdash2017}. We found that AMOEBA's individual terms are very well formulated in the asymptotic region for inherently long-ranged interactions when compared to ALMO-EDA.\cite{Mao2016,Demerdash2017,Das2018} However, there is considerable difference between the energy breakdowns from ALMO-EDA and AMOEBA on both sides of the equilibrium geometry for water and ion-water clusters. Not surprisingly, the observed difference with the ALMO-EDA energy components is likely due to the fact that most of the AMOEBA force field parameters are fitted to total energies from QM (in particular using MP2). This indirect connection to the piecewise energy terms results in an inherent inability to account for the missing CT and CP interactions to ensure cancellation of errors through their neglect. In fact we have shown that AMOEBA attempts to capture the missing CT and CP terms either through the polarization term or through its 14-7 vdW potentials by softening the VdW radii, but does so incoherently across the energy terms that depends on the particular water cluster and water-ion systems considered.\cite{Mao2016,Demerdash2017} Recently Liu et al. sought to correct this deficiency by using SAPT to guide the development of the AMOEBA+ water model whose parameters are fit to expenseive condensed phase data\cite{amoebaplus} using ForceBalance optimization\cite{Wang2013a,Wang2014}. 

In this work we describe the general and rational design of the first generation MB-UCB force field based on a complete set of decomposable functional forms, including charge transfer not available with SAPT, that captures term-by-term the ALMO-EDA decomposition of small water clusters (up through pentamers only) with no fitting to condensed phase data. The use of ALMO-EDA is motivated by its greater simplicity of the decomposed terms, avoidance of perturbation theory, the ability to separate charge transfer from polarization, and which can be combined with accurate density functionals\cite{Mardirossian2014,Mardirossian2016}. The MB-UCB force field will be the first direct test as to whether variational EDAs can offer advantages over the popular SAPT approach in practice. 

Furthermore, unlike MB-Pol, we have explicitly accounted for CP and CT that allows greater transferability to new chemistry, and also eliminating the ~1000 parameters of the high order polynomial fits that MB-Pol uses to represent the short-ranged two- and three-body terms interactions in the overlapping regime. In addition, the functional forms of the individual molecular interactions of the MB-UCB force field are designed with one primary goal in mind: the accounting of all molecular interactions that are either fit to or validated against the ALMO-EDA for greater accuracy, while maintaining a strict adherence to model choices and algorithms that keeps the computational cost manageable for condensed phase sampling, making MB-UCB much less expensive than MB-Pol. We have implemented all of the force terms for the MB-UCB model, and find that the condensed phase simulation well reproduces the radial distribution functions of the liquid as well as the temperature dependence of the density, heat of vaporization, diffusion constant, and dielectric constants, with much improved accuracy compared to previous iAMOEBA\cite{Wang2013a} and AMOEBA models\cite{Ren2003,Laury2015,amoebaplus}, and it is comparable in accuracy to MB-Pol for these same properties. This sets the stage for future reports on a complete MB-UCB potential for chain molecules such as proteins and their characterization using condensed phase simulations. 

\section{Theory}
For the direct comparison with {\it ab initio} energy components, we used the energy decomposition analysis\cite{Kitaura1976} based on the ALMO-EDA method\cite{Khaliullin2007,Horn2016} to separate the total non-bonded interaction energy into individual contributions
\begin{equation}
E_{\rm int} = E_{\rm elec} + E_{\rm Pauli} + E_{\rm disp} + E_{\rm pol} +  E_{\rm CT}
\end{equation}
where $E_{\rm elec}$, $E_{\rm Pauli}$, $ E_{\rm disp}$, $E_{\rm pol}$, and  $E_{\rm CT}$ corresponds to the contributions from the permanent electrostatics, Pauli repulsion, dispersion, polarization and charge transfer, respectively. 

Derivation of  each of the individual energy components of ALMO-EDA are described elsewhere\cite{Khaliullin2007,Horn2016}. Several points should be noted. First, the so-called geometric distortion energy is not required for our present purposes. Second, as we have discussed previously\cite{Mao2016}, the most appropriate choice for $E_{\rm elec}$ is the quasi-classical expression, which depends only on the geometry of the individual monomer of interest. Third, we use the fragment electric response function approach (at the dipole+quadrupole level) to evaluating the energy lowering due to polarization, ensuring a well-defined basis set limit.\cite{Horn2015} 

All the ALMO-EDA calculations are performed at the level of $\omega$B97X-V/def2-QZVPPD \cite{Mardirossian2014} using the Q-Chem software package\cite{Q-chem}. $\omega$B97X-V is a hybrid generalized gradient approximation (hybrid GGA) functional which includes the VV10 non-local correlation functional for dispersion correction, and has been well validated against coupled cluster benchmarks\cite{Mardirossian2014,Mardirossian2017}, including in previous assessments of AMOEBA\cite{Mao2016} The def2-QZVPPD basis\cite{Rappoport2010} has been established as very close to the basis set limit for DFT calculations of molecular interaction energies.\cite{Mardirossian2017}

The MB-UCB force field is designed for functional forms that well-reproduce the ALMO-EDA decompositions of the non-bonded terms, while the valence terms of the MB-UCB force field are the same as that used by AMOEBA\cite{Piquemal2003,Zhang2018} that in turn is based on the early MM3 force field\cite{Allinger1989}. The MB-UCB force field is subsequently validated on hexamer and large water cluster data sets at equilibrium geometries, and for highly distorted states from water dimers to decamers reported previously by Wang and co-workers\cite{Wang2014,Wang2015} . We also have  implemented the forces for the MB-UCB model and report the gOO(r), gOH(r), and gHH(r) radial distribution functions as well as a number of thermodynamic and transport properties over a range of temperatures at 1 atm. The entire model has been implemented in an in-house version of the Tinker software package\cite{tinker} and will be made available upon request.

\subsection{Permanent electrostatics}
The permanent electrostatics is modeled by considering atom centered multipoles, consisting of monopoles ($q$), dipoles ($\mu$), and with optional treatment of higher order quadrupoles $Q$. The total permanent electrostatics between all the atom pairs is expressed as

\begin{equation}
E_{\rm elec} = \sum_{i<j}{\bf M}_{i}^{ T}{\bf T}_{ij}{\bf M}_{j}
\end{equation}
where $\bf{M}$ is the multipole vector and $T_{ij}$ is the multipole interaction matrix that consists of appropriate derivatives of $\frac{1}{r_{12}}$ according to the multipole expansion. In this work we consider two multipole models in which the multipole expansion is truncated at the level of dipoles or quadrupoles on each atomic site.

Multipole expansions of the permanent electrostatics describe the anisotropy of the electrostatic interactions at mid-range and recovery of the 1/r functional form at long-range, however near equilibrium and in the very short range, when two atomic electron cloud overlaps, the multipole expansion model breaks down due to the quantum mechanical effects of charge penetration. With the exception of the GEM model\cite{Piquemal2006,Cisneros2006} which uses a density fitting formalism to express the molecular density in a Hermite Gaussian auxiliary basis, earlier incarnations of advanced force fields have used Thole damping\cite{Thole1981} functions to avoid the need to incorporate charge penetration effects. 

Charge penetration models in force fields are based on a strategy of separating the atomic charge into a core nuclear charge and smeared electron cloud charge. In fact many advanced force fields have begun to adopt some variation of the CP models\cite{Freitag2000,Piquemal2003,Piquemal2006,Cisneros2008,Slipchenko2009,Spackman2006,Wang2010,Stone2011,Tafipolsky2011,Wang2012,Rackers2017} proposed by Gordon and co-workers\cite{Slipchenko2009,Freitag2000} or Piquemal et al.\cite{Piquemal2003} The main difference between these two CP models are the use of different damping functions to approximate the value of the overlap integral. While the recent AMOEBA+ model\cite{amoebaplus} utilizes the functional form due to Gordon et al\cite{Slipchenko2009}, we have instead used the Piquemal model to account for the charge penetration effect, but only applying it as a monopole-monopole ($q-q$) correction.\cite{Wang2015} Therefore, the modified charge-charge electrostatic interactions between two atoms $A$ and $B$ with atomic charges $q_A$ and $q_B$ is expressed as
\begin{equation}\label{eq10}
E_{\rm elec}^{\rm q-q} =   \frac{Z_{\rm A}Z_{\rm B}}{\textbf {\emph r}} + \frac{Z_{\rm A}(Z_{\rm B}-q_{\rm B})}{\textbf {\emph r}}f_{\rm damp} + \frac{Z_{\rm B}(Z_{\rm A}-q_{\rm A})}{\textbf {\emph r}}f_{\rm damp} + \frac{(Z_{\rm A}-q_{\rm A})(Z_{\rm B}-q_{\rm B})}{\textbf {\emph r}}f_{\rm damp}^{\rm overlap}
\end{equation}
where, $Z$ is the effective core charge (equal to the number of valence electrons), $q$ is the atomic monopole and hence $Z-q$ describes the magnitude of the negatively charged electron cloud. The first term in Eq. \ref{eq10} is for the core-core interaction, the second and third terms describe the interaction between core and electron clouds of the other atoms and finally the fourth term accounts for the electron-electron interactions between each of the atoms. The two damping functions, $f_{\rm damp} = (1-\exp(-\alpha {\textbf {\emph r}}))$ and $f_{\rm damp}^{\rm overlap}= (1-\exp(-\beta_A \textbf {\emph r})) (1-\exp(-\beta_B \textbf {\emph r}))$, require two parameters, $\alpha$ and $\beta$, to control the damping of core-electron and electron-electron interactions, respectively. From Eq. \ref{eq10} and the two damping functions, it is clear that the charge penetration corrections can be made to decrease rapidly, and that the electrostatic energy will correctly reduce to classical Coulombic multipolar interactions in the medium and asymptotic long-range limits. 

\subsection{Polarization energy}
Polarization is explicitly incorporated in MB-UCB by induced dipoles at  polarizable sites located on the atomic centers\cite{Ren2003}. The induced dipoles ($\boldsymbol\mu^{\rm ind}$) at a polarizable site $i$ with polarizability $\alpha_i$ is expressed as
\begin{equation}\label{eq3}
\boldsymbol\mu^{\rm ind}_{i} = \alpha_i \left[ \sum_{j} {\bf{T}}_{ij} {\bf{M}}_j - \sum_{j\neq i} {\bf{T}}^{d-d}_{ij}\boldsymbol\mu^{\rm ind}_j\right]
\end{equation}
Where $\bf{T}$ and $\bf{M}$ are the multipole-multipole interaction matrix and polytensor permanent multipoles, respectively, and $\bf{T}^{d-d}$ is the  dipole-dipole interaction tensor. The first term of Eq. \ref{eq3} is for the {\it direct} polarization of the induced dipoles by the permanent multipoles with field ${\bf E}_i$ and the second term is the {\it mutual} induction by induced dipoles on all the other sites. Thus Eq. \ref{eq3} can be rewritten as

\begin{equation}\label{eq4}
\boldsymbol\mu^{\rm ind}_{i}  = \alpha_i \left[{\bf{E_i}} - \sum_{j\neq i} {\bf{T}}^{d-d}_{ij}\boldsymbol\mu^{\rm ind}_j \right]
\end{equation}

\begin{equation}\label{eq5}
\implyarrow \alpha_i^{-1} \boldsymbol\mu^{\rm ind}_{i} + \sum_{j\neq i} {\bf{T}}^{d-d}_{ij}\boldsymbol\mu^{\rm ind}_j  = \bf{E_i}
\end{equation}
and can be represented in general matrix form as 
   \begin{equation}\label{eq6}
     \begin{pmatrix}
    \alpha_{1}^{-1} & T_{1\,2}^{d-d} & \cdots  & T_{1\,N}^{d-d} \\
     T_{1\,2}^{d-d} & \alpha_{2}^{-1}  & \cdots  & T_{2\,N}^{d-d}   \\
    \vdots   & \vdots   & \ddots  & \vdots   \\
     T_{1\,N}^{d-d} & T_{2\,N}^{d-d}  & \cdots  & \alpha_{N}^{-1} 
     \end{pmatrix}
     \begin{pmatrix}
       \mu_{1} \\
       \mu_{2} \\
      \vdots \\
       \mu_{N}
     \end{pmatrix}
     =
     \begin{pmatrix}
       E_{1} \\
       E_{2} \\
       \vdots \\
       E_{N} \\
     \end{pmatrix}
   \end{equation}

In Eq.~\ref{eq6}, the off-diagonal blocks $\bf{T}^{d-d}$ are the Thole damped\citep{Thole1981} Cartesian interaction tensors between induced dipoles of two polarizable sites $i$ and $j$, while the diagonal blocks are the inverse of the atomic polarizability.

Because in an anisotropic media an applied electric field will induce an anisotropic polarization response, we recently introduced the model for the anisotropic atomic polarizabilities\cite{Das2018} as a rank two tensor. 

\begin{equation}\label{eq7}
\alpha^{-1}_i= 
\begin{pmatrix}
\alpha_{i,xx} & \alpha_{i,xy}  & \alpha_{i,xz} \\
\alpha_{i,yx}  & \alpha_{i,yy}  & \alpha_{i,yz}  \\
\alpha_{i,zx}  & \alpha_{i,zy}  & \alpha_{i,zz}  
\end{pmatrix}^{\!-1}
\end{equation}
Our use of anisotropic polarization differs from the standard AMOEBA and AMOEBA+ models which assumes isotropic polarization\cite{Wang2013a,Laury2015,amoebaplus}. 

In this work we use a standard conjugate gradient self-consistent field (CG-SCF) solver\cite{Wang2005} to obtain the induced dipoles and the polarization energy ($E_{\rm pol}$) using Eqs.~\ref{eq6} and \ref{eq7} 
\begin{equation}\label{eq8}
E_{\rm pol} = -\frac{1}{2}\sum_i\\\boldsymbol\mu_i^{\rm ind}\bf{E}_i
\end{equation}
In future work the anisotropic polarization will be combined with our iEL/SCF and SCF-less approaches that reduces the computational expense of this step\cite{Albaugh2015,Albaugh2017}, while also adding the benefits of a resonance controlled  multi-time stepping algorithm\cite{Albaugh2019}, when used in molecular dynamics.

\subsection{Charge transfer energy}

Recently, Deng et al.\cite{Shideng2017} used an empirical many-body function to decompose SAPT induction energy into polarization  and charge transfer energies. Here we focus on their CT expression, given by 
\begin{equation}
\label{eq13}
 E_{\rm CT} = -\frac{1}{2}\sum_i \boldsymbol\mu_i^{\rm ct} \bf{E}_i^{\rm ct}
\end{equation}
\begin{equation}
\boldsymbol\mu^{\rm ct}_{i} = \alpha_i^{\rm ct} \sum_{j} {\bf{T}}_{ij}^{\rm ct} {\bf{M}}_j
\end{equation}
where $\alpha_i^{\rm ct}$ controls the charge transfer energy between two atoms through a response to the permanent electrostatics. To be clear, we note that this CT model has no explicit charge flow. The multipole interaction matrix ($\bf{T}^{\rm ct}$) elements are damped with an exponential damping function 

\begin{equation}
{\bf T}_{\zeta}^{\rm ct} = -f_3^{\rm ct} \frac{r_\zeta}{r_{ij}^3}, \zeta = x, y, z
\end{equation}
where
\begin{equation}
f_3^{\rm ct} = [1-{\rm d}\exp(-{\rm b}u^3)], u = \frac{r_{ij}}{(\alpha^{\rm ct}_i \alpha^{\rm ct}_j )^{\frac{1}{6}}}
\end{equation}
in which the two parameters $b$ and $d$ are responsible for the exponential decay of the charge transfer energy, which should be more short-ranged that polarization. However, unlike the Deng et al. model\cite{Shideng2017}, which only considered the direct charge transfer between the atomic sites, we also consider the mutual CT term
\begin{equation}
\label{eq15}
 E_{\rm CT-ind} = -\frac{1}{2}\sum_i \boldsymbol\mu_i^{\rm ct-ind} \bf{E}_i^{\rm ct}
\end{equation}
\begin{equation}
{\boldsymbol \mu}^{\rm ct-ind}_{i} = \alpha_i^{\rm ct} \left[ \sum_{j} {\bf T}_{ij}^{\rm ct} {\bf M}_j - \sum_{j\neq i} {\bf T}^{{\rm ct}{\rm   [d-d]}}_{ij}{\boldsymbol \mu}^{\rm ct-ind}_j\right]
\end{equation}
that recovers a greater amount of the many-body character of charge transfer, albeit with the understanding that no explicit charge flow is operative. By contrast the AMOEBA+ model includes only pairwise charge transfer, and furthermore must assume some arbitrary amount of charge transfer since it is not defined explicitly in SAPT.\cite{amoebaplus}

\subsection{vdW interactions}
The remaining energy terms in ALMO-EDA, Pauli repulsion and dispersion, are modeled in MB-UCB as a van der Waals interaction using a  buffered 14-7 pairwise-additive function proposed by Halgren\cite{Halgren}
\begin{equation}\label{eqvdw}
E_{\rm vdW} = \sum_{i<j}\epsilon_{ij}\left( \frac{1+\delta}{\sigma_{ij}+\delta}\right)^7\left( \frac{1+\gamma}{\sigma_{ij}^7+\gamma} - 2\right)
\end{equation}
$\epsilon$ defines the energy scale, and $\sigma = \frac{r}{R_0}$ is the dimensionless distance between two atoms, where $R_0$ is the distance corresponding to the  minimum energy. Like AMOEBA\cite{Ponder200327}, we set the two constants $\delta$ and $\gamma$ to 0.12 and 0.07, respectively, while $\epsilon$ is optimized for the MB-UCB model. 

\subsection{Parameterization Strategy}
In previous studies of the many-body expansion of the AMOEBA polarizable model, we found that liquid water potential energies were fully recovered through four-body terms, while truncation at the level of five-body terms were required to reproduce forces.\cite{Demerdash2016} Hence the MB-UCB model is parameterized on the water monomer, dimer, trimer, tetramer, and pentamer cluster data sets only. 

In our previous study of polarization anisotropy, the atomic permanent electrostatics and anisotropic polarizabilities were obtained from the {\it ab initio} calculations at the level of  $\omega$B97X-V/def2-QZVPP using the Williams-Stone-Misquitta (WSM) generalized distributed multipole analysis (GDMA)\cite{Misquitta2006} with the help of CamCASP suite program\cite{camcasp}. We kept the same monopole and dipole parameters as the original AMOEBA03 model\cite{Ren2003}, and optimized the atomic quadrupoles to reproduce the {\it ab initio} electrostatic potential (ESP) (using MP2/aug-cc-pVTZ) on a grid of points outside the van der Waals surface (MB-UCB-MDQ).  

However in this work we have made a different design decision by eliminating the permanent quadrupoles, and thus we have again used the WSM approach to reoptimize both the permanent monopole and dipole parameters with anisotropic polarization to reproduce the single water molecule electrostatics (MB-UCB-MD). To improve the electrostatics in the short range further, we have adopted the charge penetration model with parameters taken directly from the literature\cite{Wang2015}, but only applied to the monopole electrostatics. We later present results establishing that there is no need to refit their CP parameters for either the MB-UCB-MDQ or MB-UCB-MD permanent electrostatics. We note that we do not fit to ALMO-EDA for the permanent electrostatics or polarization, but use it simply as a validation tool for the geometric scans of small water clusters. But for the  charge transfer model, the CT parameters were fitted to reproduce the ALMO-EDA charge transfer energies for the data set consisting of water dimers, trimers, tetramers and pentamers extracted from AMOEBA MD simulations. 

Finally for the van der Waals interaction of the MB-UCB model, we fitted the parameters in Eq. \ref{eq17} from the remaining energy difference between the $\omega$B97X-V/def2-QZVPPD total binding energy and the previous model energy terms 
\begin{equation}
E_{\rm Pauli} + E_{\rm disp} = E_{\rm  abinito} - E_{\rm elec} -  E_{\rm pol} -  E_{\rm CT} - E_{\rm valence}
\label{eq17}
\end{equation}
that will not only account for the Pauli and dispersion interactions, but will also clean up any disparities between ALMO-EDA and the individual $E_{\rm elec}$, $E_{\rm pol}$, and $E_{\rm CT}$ terms of MB-UCB. 

The van der Waals parameters were optimized using data sets containing more than 9604 geometries consisting of water dimers, trimers, tetramers, and pentamers. Then, we performed a validation study of our complete MB-UCB model over larger water cluster data sets and liquid water properties. 

\subsection{Simulation Protocol}
Condensed phase properties were characterized in in the isothermal-isobaric (NpT) ensemble ensemble for 214 water molecules at 1 atm and over a range of temperatures. The equations of motion were integrated using the velocity Verlet algorithm\cite{Swope1982}  and with a time step size of 1 femtosecond. The duration of equilibration runs was 500 picoseconds and production runs were 4 nanoseconds.   Temperature and pressure were controlled using Nose Hoover methods\cite{Martyna1996} and Coulomb interactions were computed using Ewald summation. For the diffusion constants, a collection of 50 independent snapshots were used as starting points for separate NVE trajectories of 100ps length each. Using the molecular dynamics simulations we have calculated the radial distribution functions and dielectric constant at room temperature, and the temperature dependence of the density, heat of vaporization, heat capacity, and diffusion constants (which have been corrected for finite size effects), using previously reported protocols.\cite{Horn2004,Wang2013a}

\section{Results}
For consideration of the breakdown of individual intermolecular interactions, we performed a piecewise decomposition of the MB-UCB force field to directly compare with the ALMO-EDA components for the water dimer and trimer along various scanned coordinates. For the water dimer, the oxygen-oxygen  distance is used as the scanned coordinate, where as for water trimer one of the water molecules is moved from the centroid of the triangle formed by the three oxygen atoms.\cite{Mao2016,Demerdash2017} Additionally, permanent electrostatics were computed for 50 random configurations extracted from a 50 ps MD simulation  at 298 K, with each configuration separated by 1 ps.

\begin{figure}[H]
  \includegraphics[scale=0.07]{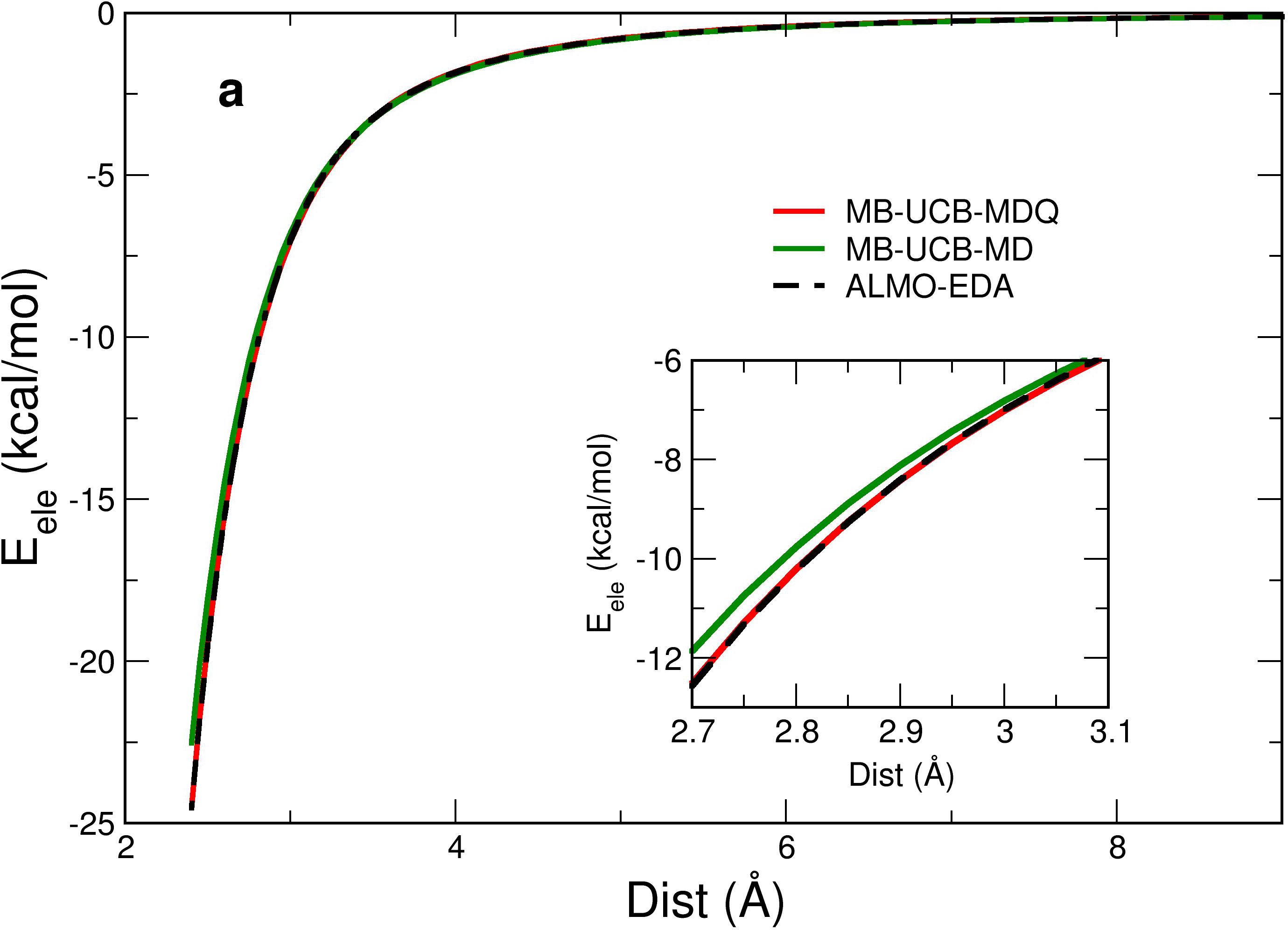}
 \includegraphics[scale=0.07]{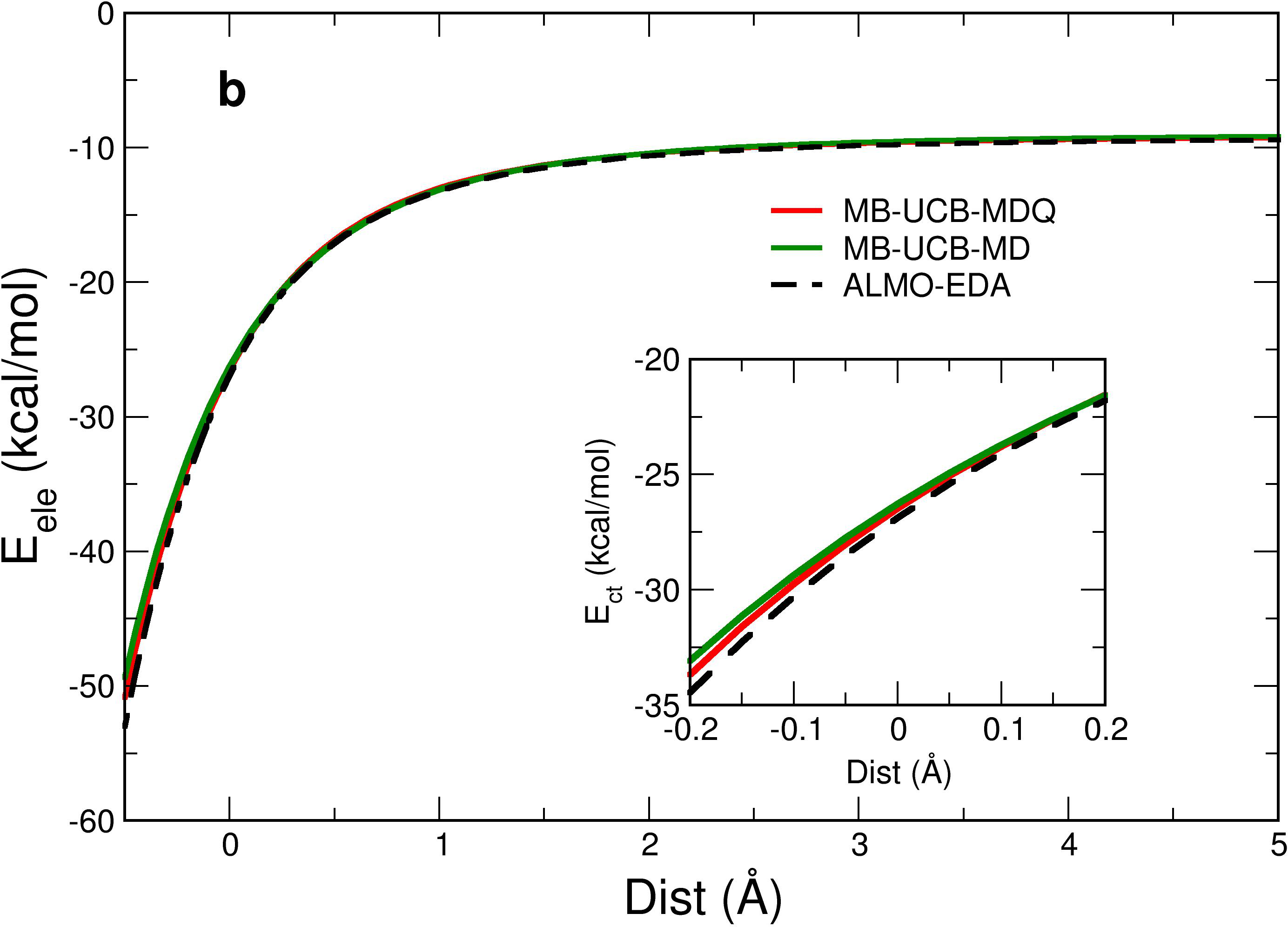}
  \includegraphics[scale=0.07]{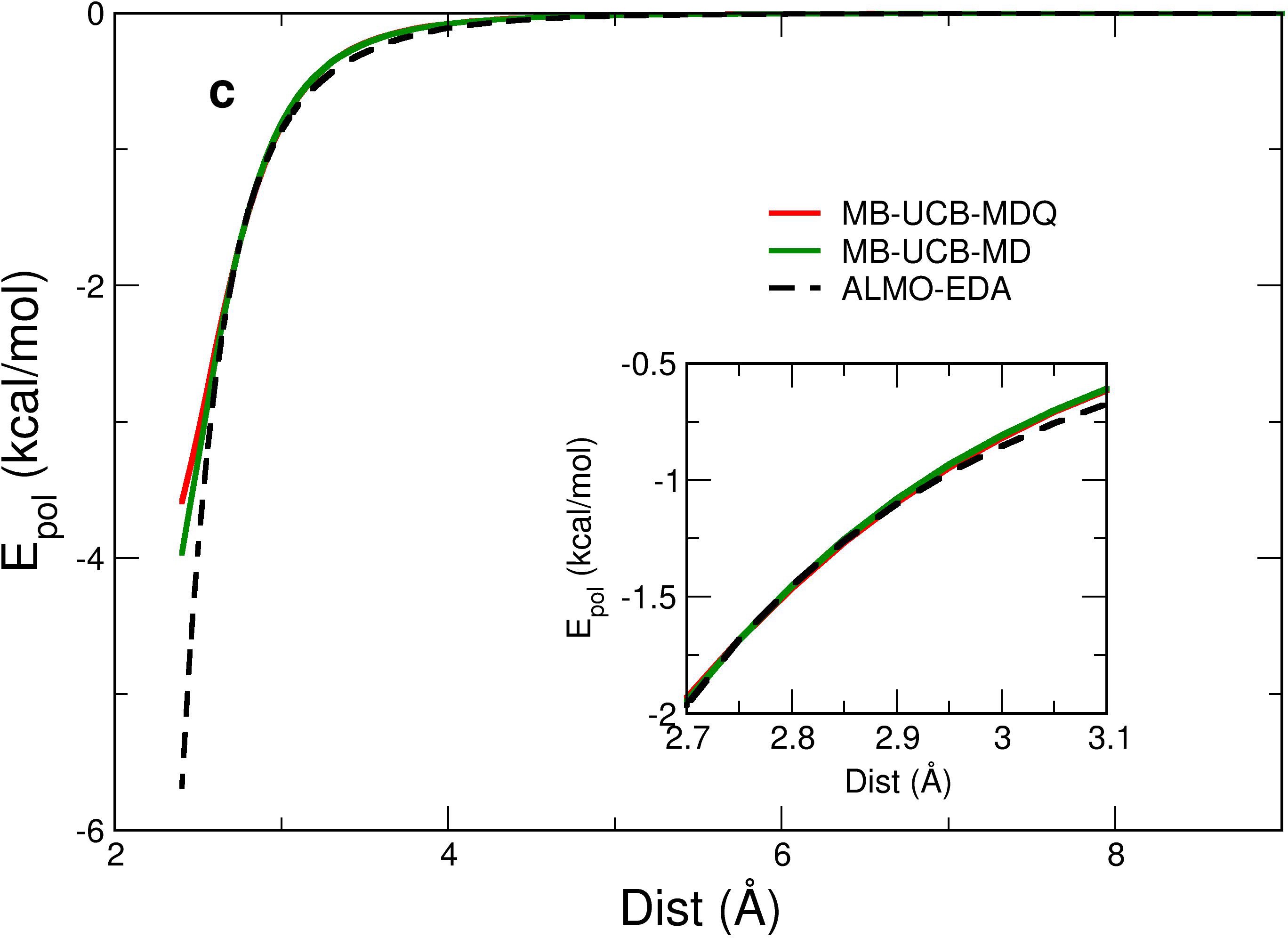}
 \includegraphics[scale=0.07]{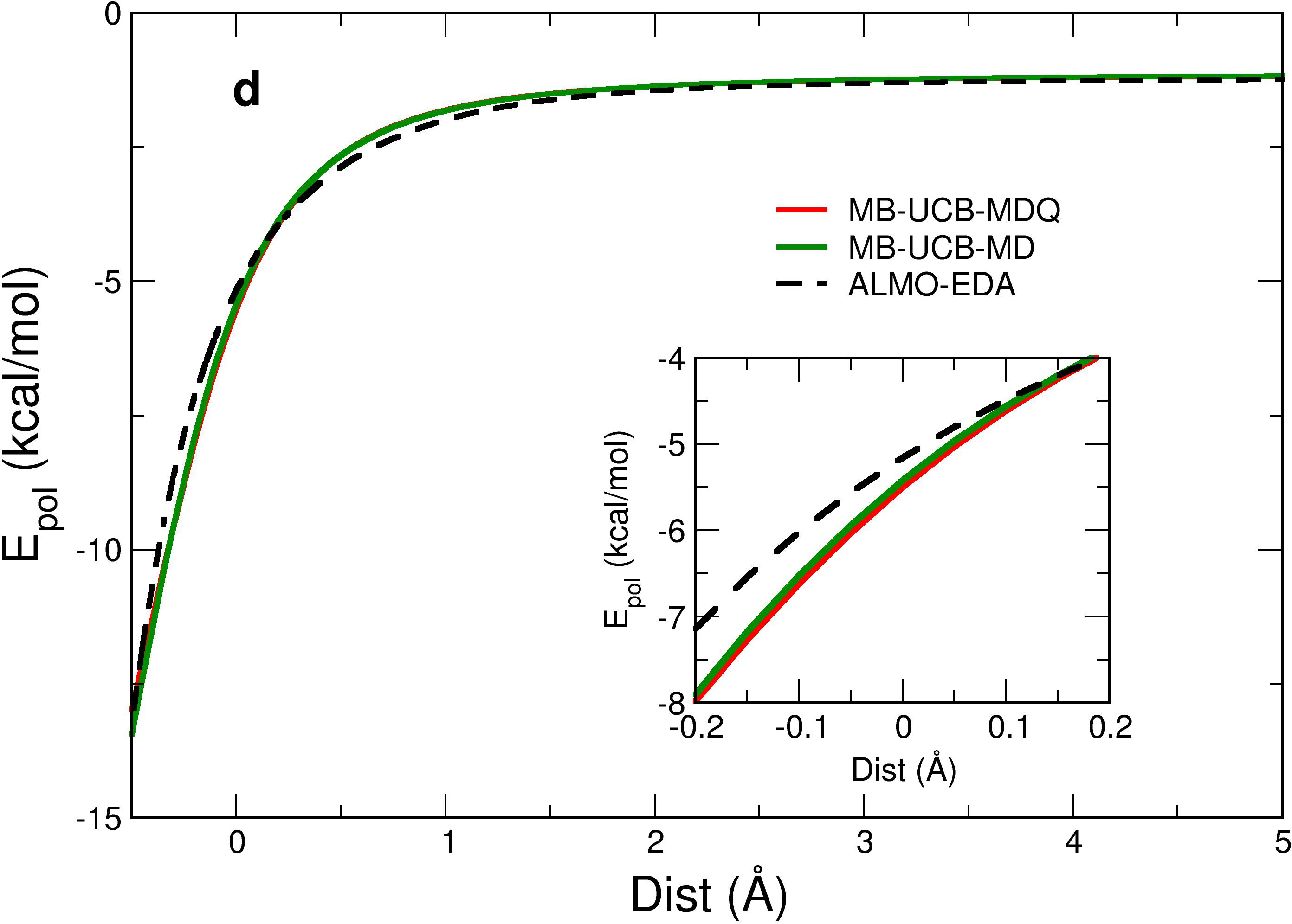}  \includegraphics[scale=0.07]{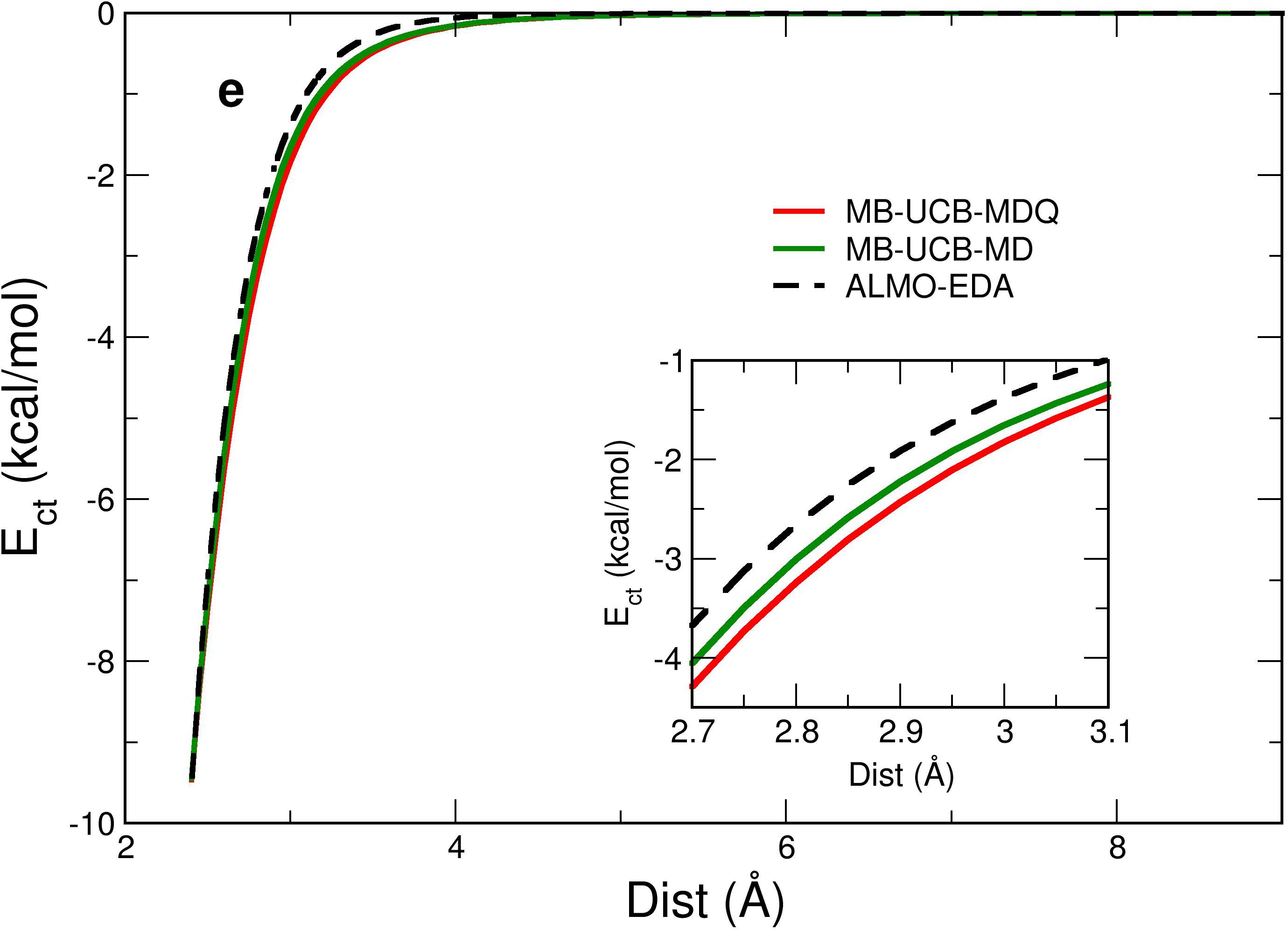}
 \includegraphics[scale=0.07]{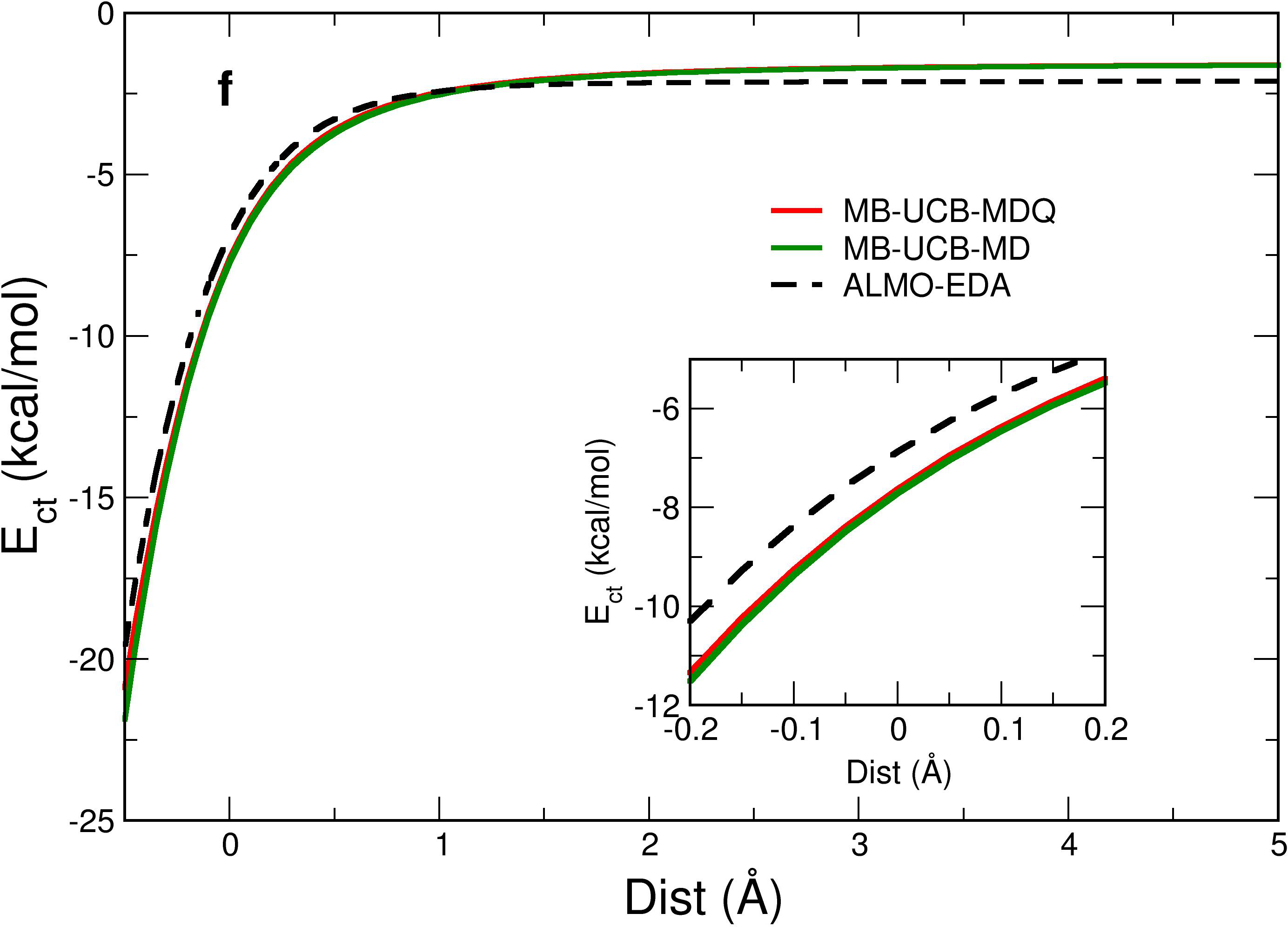}\caption{Change in (a,b) permanent electrostatics, (c,d) polarization energy, and (e,f) charge transfer energy for the water dimer (H$_2$O)$_2$ (left) and water trimer (H$_2$O)$_3$ (right) along the scanned coordinate for ALMO-EDA and the MB-UCB models. The 0 \AA\/ distance in the water trimer corresponds to the equilibrium geometry. We consider truncation of the multipole expansion of the MB-UCB models up to dipoles (MD) and up to quadrupoles (MDQ)}
 \label{fig:fig1}
 \end{figure}

From Figures \ref{fig:fig1}a and \ref{fig:fig1}b it is clear that for the permanent electrostatics with inclusion of CP that the MB-UCB model is in excellent agreement with the ALMO-EDA result not only asymptotically but in the compressed region as well. It should be emphasized that we do not fit the permanent electrostatics to ALMO-EDA, and hence this provides clear evidence of compatibility between $E_{\rm elec}$ from ALMO-EDA and the CP-corrected electrostatics. An equally important result is that while the MDQ multipole truncation of the MB-UCB force field agrees very well with the ALMO-EDA result, there is only small degradation for MB-UCB-MD in which the energy differences are  within chemical accuracy of $<$ 1 kcal/mole.

\begin{figure}[H]
\includegraphics[scale=0.07]{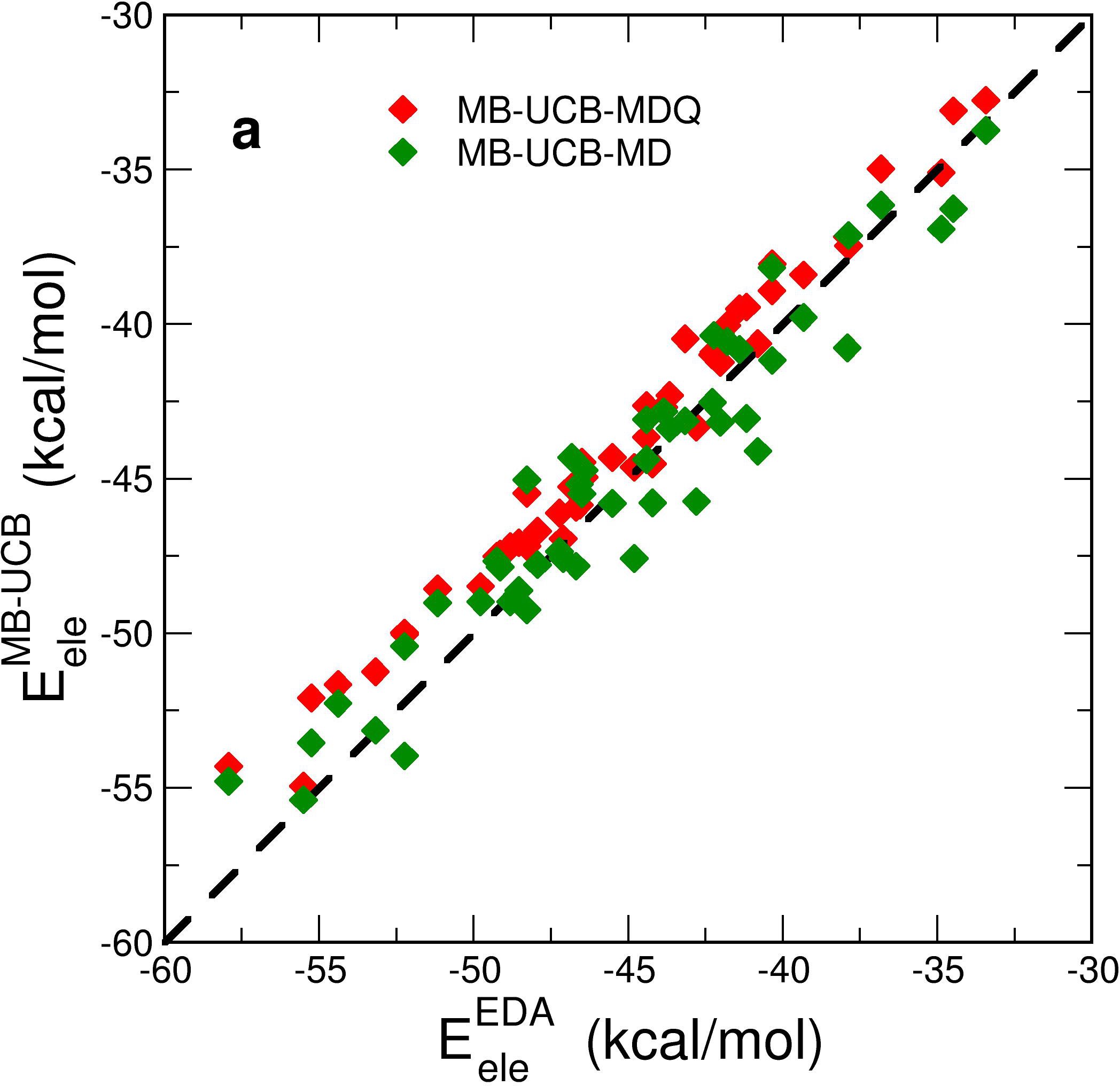}
\includegraphics[scale=0.07]{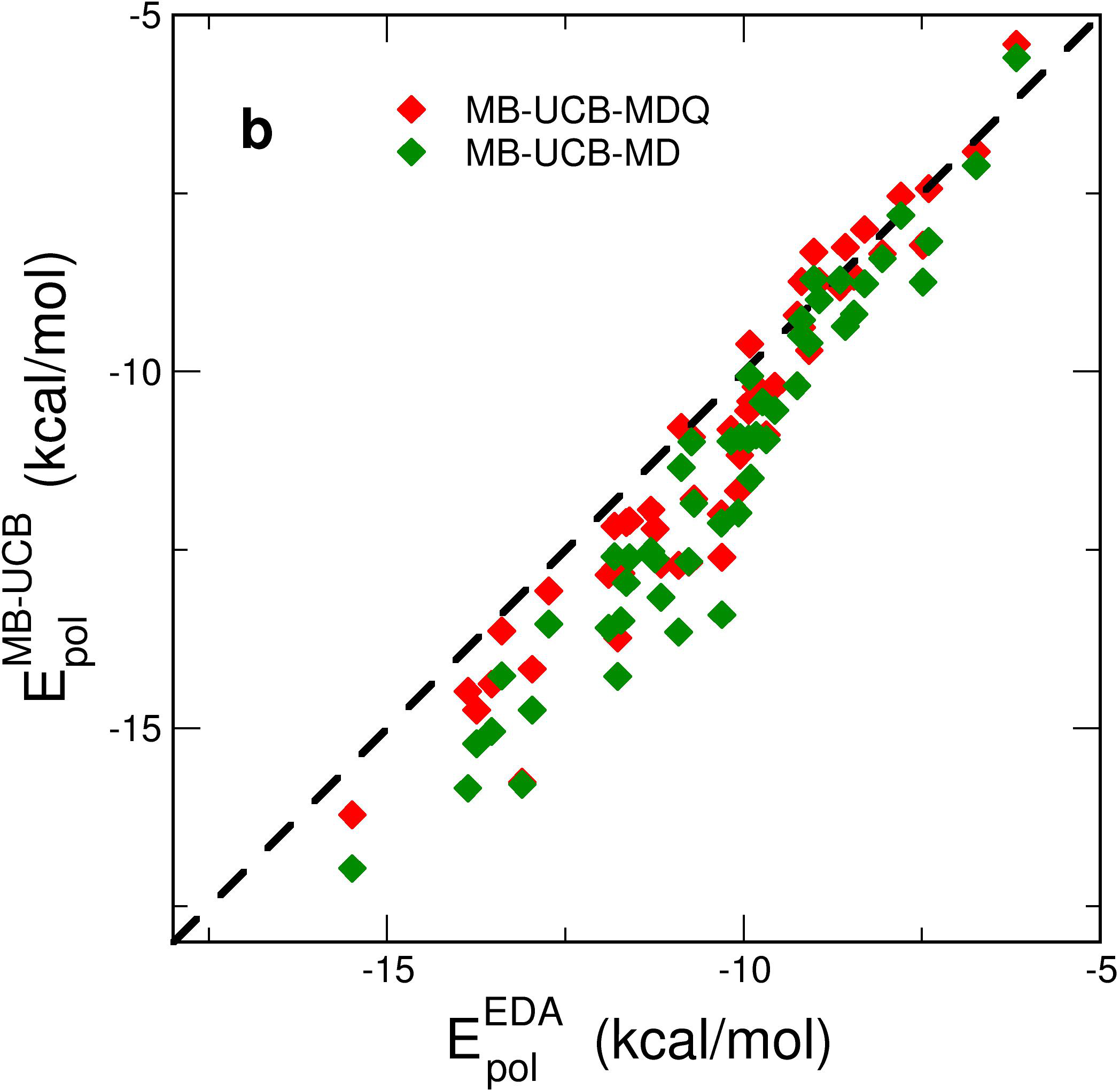}  
\includegraphics[scale=0.07]{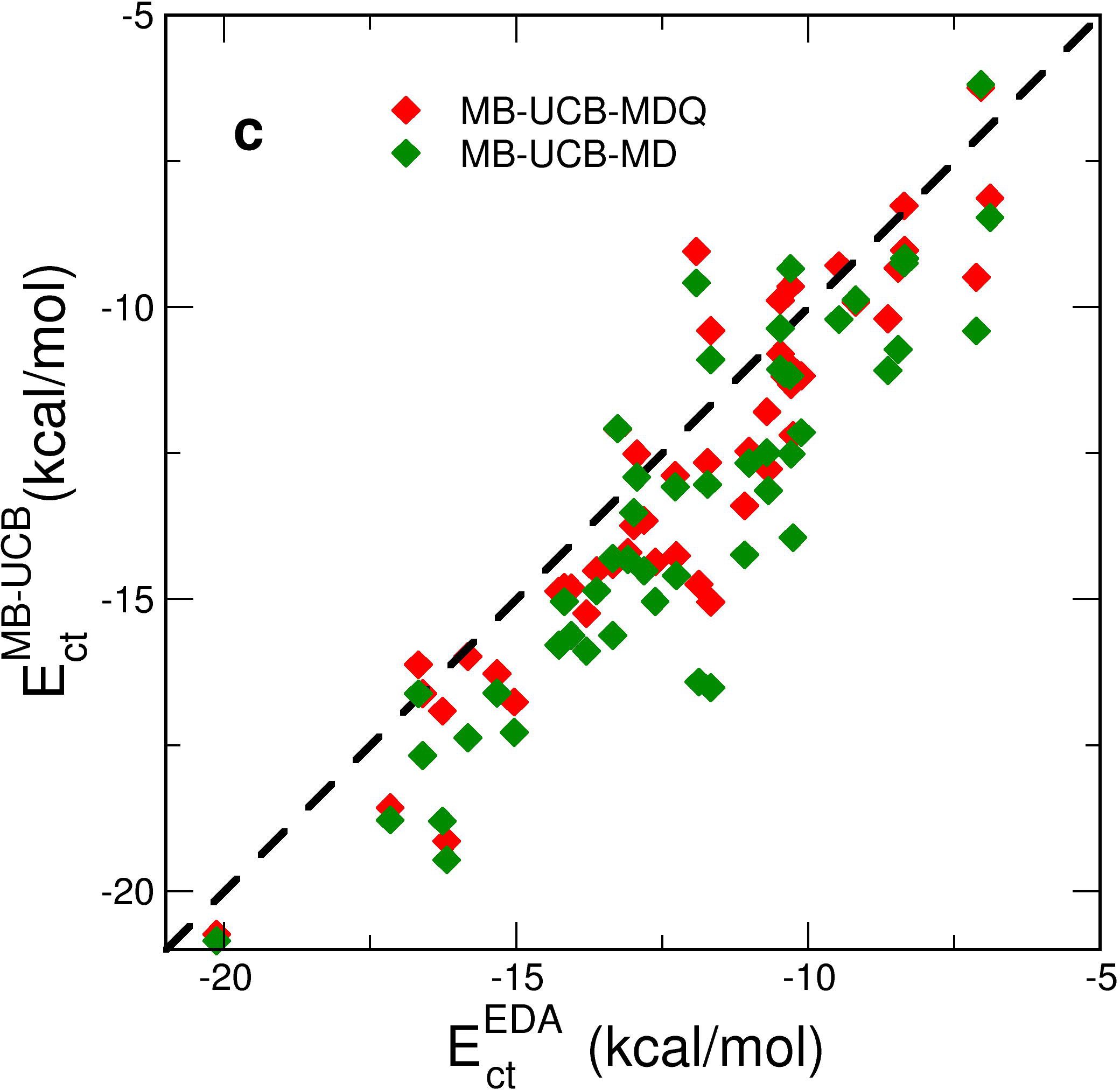}
\caption{Correlation of (a) electrostatic, (b) polarization, and (c) charge transfer energies obtained from ALMO-EDA and MB-UCB using monopole-dipole (MD) and monopole-dipole-quadrupole (MDQ) for 50 random configurations extracted from a MD simulation.}
 \label{fig:fig2}
 \end{figure}

Figure \ref{fig:fig2}a  further supports the conclusion that the permanent electrostatics are in good agreement between ALMO-EDA and MB-UCB when examining the water pentamer set derived from a molecular dynamics trajectory. Again we find very little sensitivity to the truncation level of the multipole expansion, suggesting that the monopole-dipole electrostatic design choice will be a good one, at least for water.

We have previously shown that the incorporation of atomic anisotropic polarizabilities in the MB-UCB offers significantly better agreement with the ALMO-EDA polarization for the water dimer, trimer and the MD-pentamer results than found using standard AMOEBA03.\cite{Ren2003} Figure \ref{fig:fig1}c  re-examines the anisotropic polarization energy of MB-UCB for the water dimer, but this time for the two multipole truncations, in which we find that both agree very well with the ALMO-EDA polarization energy.\cite{Horn2015} Again the anisotropic polarization of the MB-UCB model was developed independently of the ALMO-EDA result, suggesting that using two very different approaches for deriving the fixed electrostatics and the leading 2-body polarization response, that the GDMA and ALMO-EDA decompositions mutually reinforce their basic correctness. However for the water trimer and the distorted pentamers shown in \ref{fig:fig1}d and \ref{fig:fig2}b, the MB-UCB model exhibits some over-polarization in comparison to ALMO-EDA, although much less so than what we originally found for AMOEBA03 as reported in previous work.\cite{Das2018} We do note that the fragment electric response function utilized by the ALMO-EDA has a contribution from Pauli repulsion (through electron reconfiguration) that contributes to energy lowering that would not explain the differences. Hence these energetic discrepancies must be compensated for in the last stage of the van der Waals parameterization.

The MB-UCB model differs from most other many-body force fields with the inclusion of a model for the charge transfer interaction. Using a parameterization that is intentionally fit to the ALMO-EDA decomposition for charge transfer, the MB-UCB force field's computed charge transfer energies are thus compared between the two multipole truncations for both the water dimer and water trimer in Figures \ref{fig:fig1}e, \ref{fig:fig1}f, and \ref{fig:fig2}c. It is evident that the general induction functional form used for the charge transfer energy is able to reasonably recapitulate the {\it ab initio} charge transfer energy from ALMO-EDA using either level of multipole truncation, but with less fidelity than found for the electrostatics. It is certainly desirable to reconsider the choice of functional form in future work to better describe charge flow.  Even so, the level of qualitative and even quantitative agreement is quite good given the limitations of the functional form.

\begin{figure}[H]
  \includegraphics[scale=0.07]{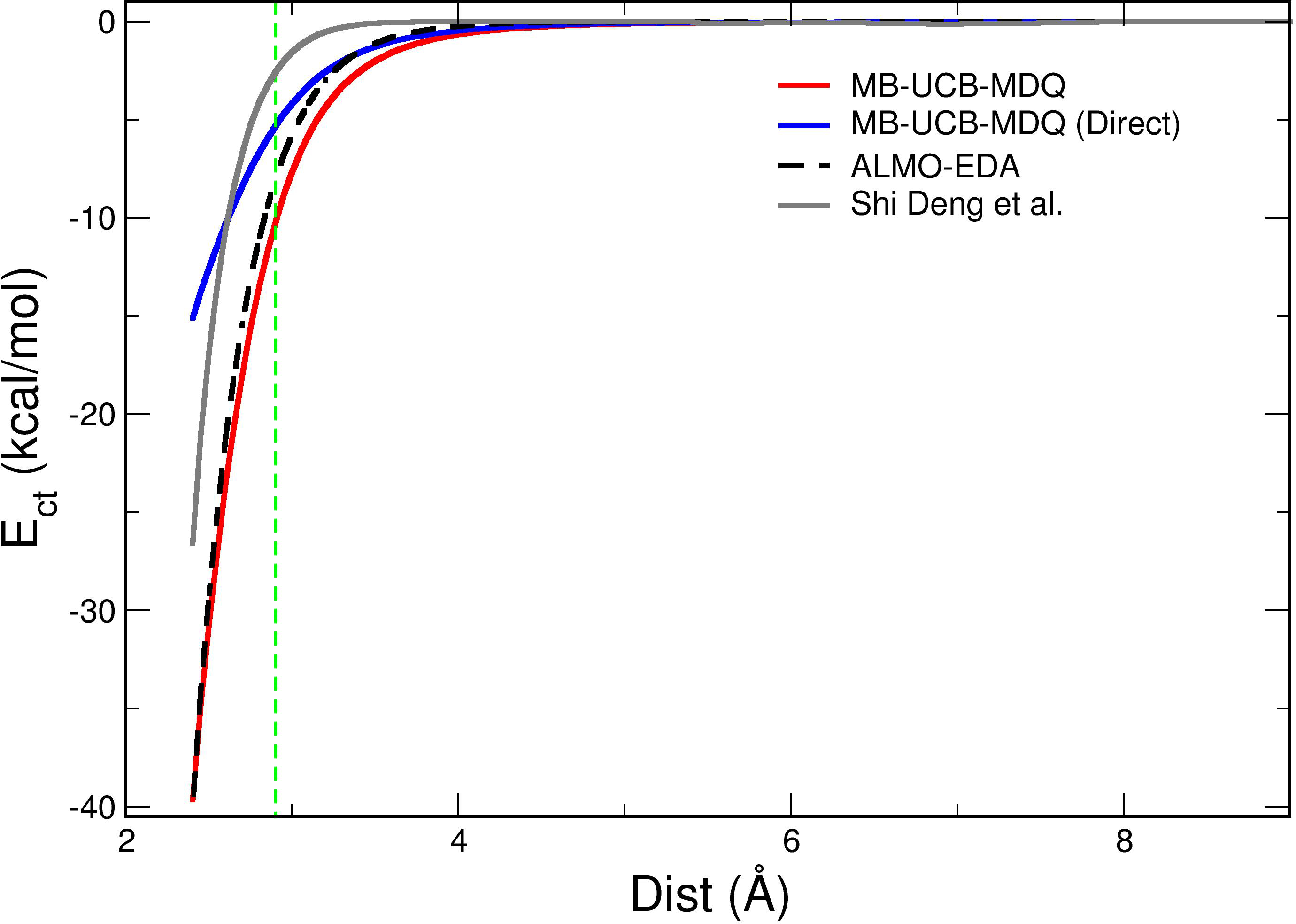}
 \includegraphics[scale=0.07]{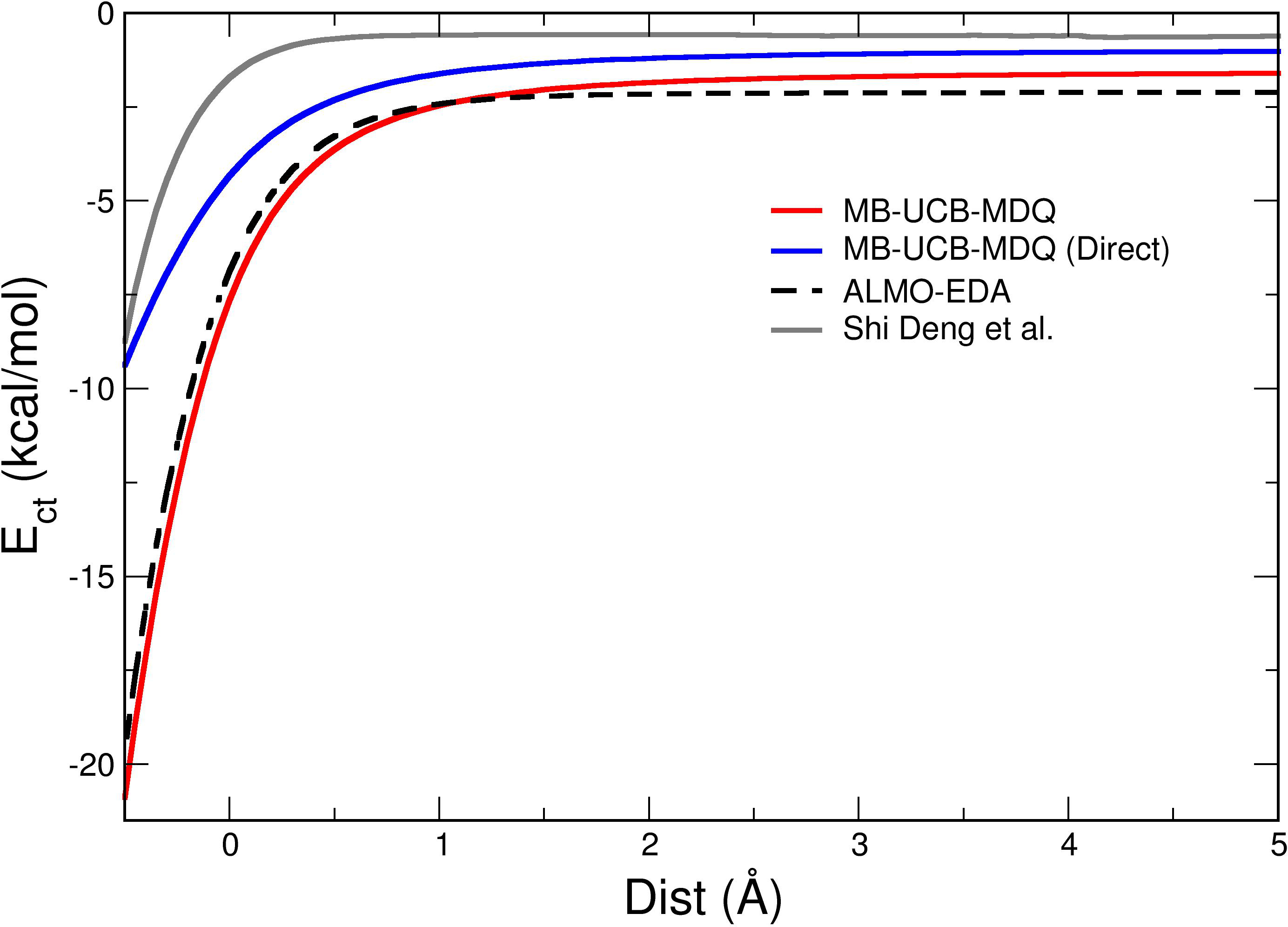}
 \caption{Different components of the charge transfer energy along the scanned coordinate for water dimer (H$_2$O)$_2$  (left) and trimer (H$_2$O)$_3$ (right) using decomposition of SAPT vs ALMO-EDA. Dotted green line is the equilibrium O-O distance in water dimer. Distance 0 \AA\/ in the water trimer corresponds to the equilibrium geometry.}
 \label{fig:fig3}
 \end{figure}
 
It is worth considering the charge transfer energy calculated using the SAPT decomposition proposed by Deng et al.\cite{Shideng2017}, which is seen to underestimate the charge transfer energy compared to the ALMO-EDA for the water dimer and trimer (Figure \ref{fig:fig3}). The underestimation of their CT energy compared to MB-UCB and ALMO-EDA can be due to either the neglect of the mutual charge transfer energy in their induction formulation, and/or their empirical separation of the SAPT induction energy into polarization and charge transfer components. We can test one of these two possibilities directly by removing the mutual charge transfer operation in the MB-UCB model, leaving behind only the direct charge transfer energy, showing that indeed the charge transfer energy would be underestimated in the short range as a result (see Figure \ref{fig:fig3}a). Given that the agreement between ALMO-EDA and MB-UCB model using GDMA for the polarization energy is mutually reinforcing, i.e. given that they agree well although the intermolecular polarization is derived independently of each other, we conclude that the empirical separation of the SAPT induction energy has underestimated the charge transfer and overestimated the polarization energy\cite{Shideng2017}. This conclusion is consistent with detailed comparison of the energy decomposition schemes themselves on a variety of model systems.\cite{Mao2018} A more detailed comparison of different energy components for the water dimer obtained from energy decoposition analysis using SAPT2+\cite{Hohenstein2011} (SAPT2+/def2-QZVPPD) and variational ALMO-EDA ($\omega$B97X-V/def2-QZVPPD) is shown in Figure S1.

\begin{figure}[H]
  \includegraphics[scale=0.07]{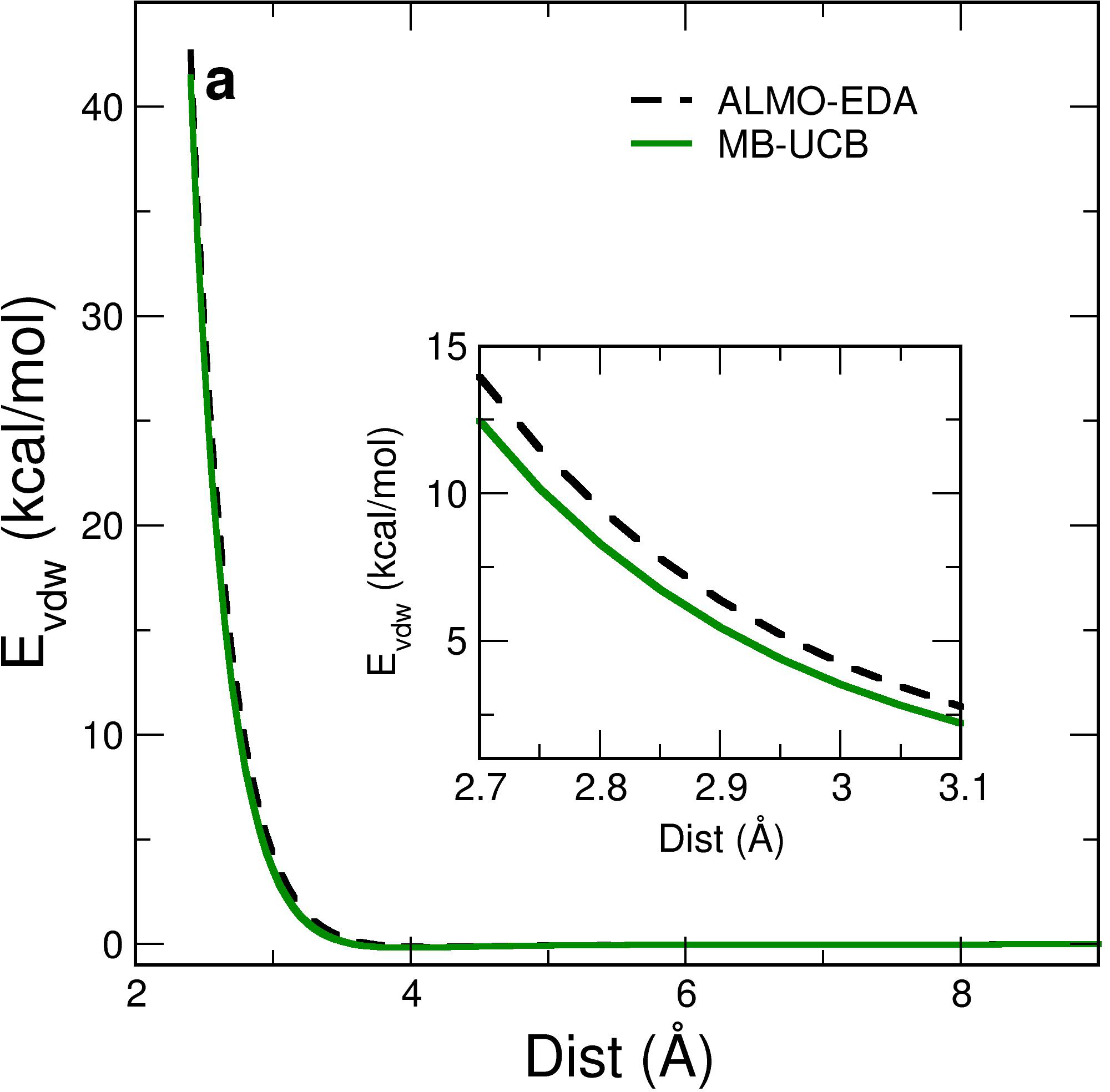}
 \includegraphics[scale=0.07]{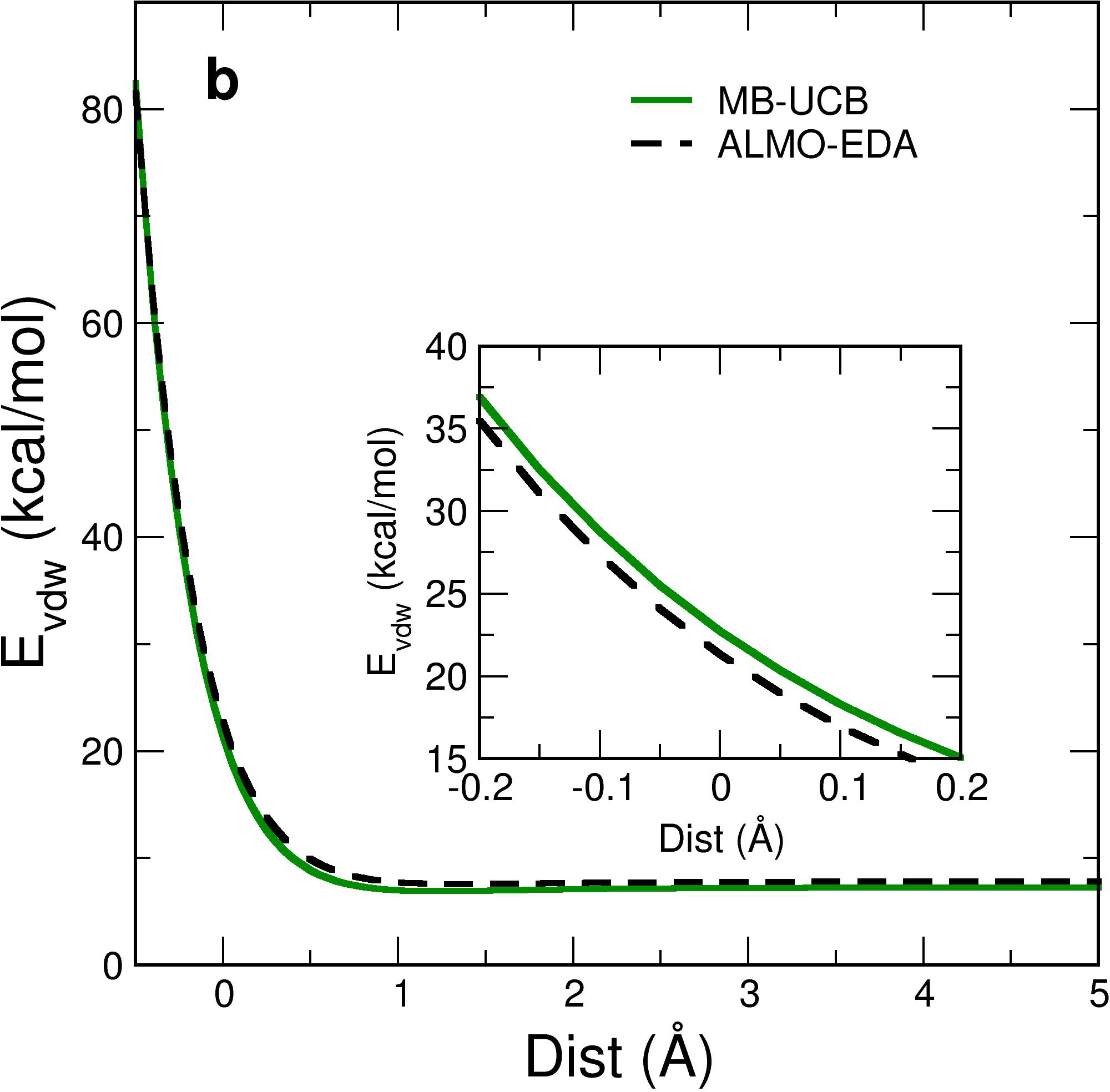}
 \includegraphics[scale=0.07]{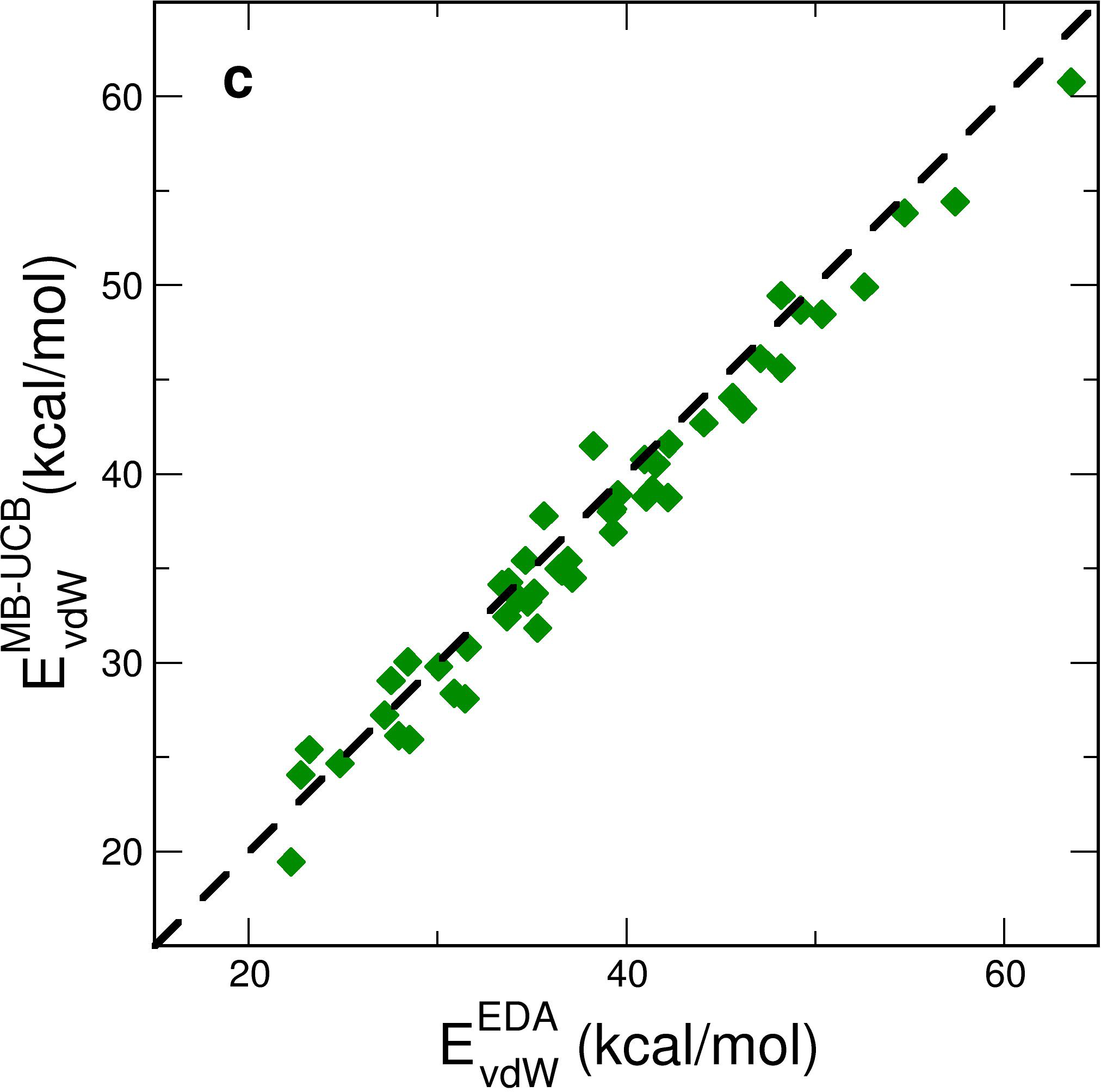}
 \caption{Change in the van der Waals energy along the scanned coordinate for water dimer (H$_2$O)$_2$  (a), trimer (H$_2$O)$_3$ (b) and the correlation of vdW energy between ALMO-EDA and MB-UCB of 50 random configurations extracted from a MD simulation (c). Distance 0 \AA\/ in the water trimer corresponds to the equilibrium geometry.}
 \label{fig:fig4}
 \end{figure}
 
In Figure \ref{fig:fig4} we compare the van der Waals energy of the MB-UCB force field using Eq. \ref{eqvdw} against the combined energy term E$_{\rm Pauli}$ + E$_{\rm disp}$ from ALMO-EDA. For both the water dimer and trimer, the MB-UCB van der Waals energy agrees quite well with the ALMO-EDA through out the scanned coordinates, with a similar trend observed for the MD extracted water pentamers. What is pleasing about this result is that the MB-UCB model is also fitting any  disagreements between the individual $E_{\rm elec}$, $E_{\rm pol}$, and $E_{\rm CT}$ terms of MB-UCB with ALMO-EDA. Thus the excellent agreement in the van der Waals energy is a reflection that there is very effective cancellation of small errors among these terms.

Table 1 provides a validation suite for dimers through 20-mer water cluster conformational energies at stationary point geometries (minima and transition states) evaluated with MP2 or CCSD(T) taken from previously reported literature\cite{van2003,Ren2003,Fanourgakis2004,Xantheas2004,Bulusu2006,Bates2009,Yoo2010}. Overall the MB-UCB-MDQ and MB-UCB-MD models are in good agreement with the reference ab initio data for small water clusters, with larger errors for larger clusters, but all are within the same range of error of AMOEBA\cite{Ren2003}, AMOEBA14\cite{Laury2015} or iAMOEBA\cite{Wang2013a} force fields.

\begin{table}[H]
\caption{Comparison of binding energies between MB-UCB and  {\it ab initio} references for different  water clusters of various sizes. }
\resizebox{\textwidth}{!}{%
\begin{tabular}{l l c c c c c c c}
\hline
  & Molecule & Previous Benchmarks & DFT($\omega$B97X-V/def2-QZVPPD)\cite{Mardirossian2014} & MB-UCB-MDQ & MB-UCB-MD \\
  \hline
Dimers\cite{van2003}(Smith) & 1      &   -4.968  & -4.87188    & -5.1540    &  -5.0596   \\
 & 2                       &   -4.453  & -4.34329    & -4.7759    &  -5.0090   \\
 & 3                       &   -4.418  & -4.30378    & -3.8606    &  -4.0418   \\
 & 4                       &   -4.25   & -4.05396    & -3.1060    &  -3.6774   \\
 & 5                       &   -3.998  & -3.77482    & -3.6831    &  -3.4716   \\
 & 6                       &   -3.957  & -3.71311    & -3.2071    &  -3.5344   \\
 & 7                       &   -3.256  & -3.20493    & -2.9349    &  -2.5900   \\
 & 8                       &   -1.3    & -1.38317    & -1.1501    &  -0.8613   \\
 & 9                       &   -3.047  & -3.04798    & -2.9863    &  -2.6375   \\
 & 10                      &   -2.182  & -2.22805    & -2.0673    &  -2.0865   \\
Trimer\cite{Ren2003} &  & -15.742 & -15.7235    & -16.0730    & -16.6938    \\
Tetramer\cite{Ren2003} &           &   -27.4   & -27.6986    & -27.8782    & -28.0060    \\
Pentamer\cite{Ren2003} &    &   -35.933 & -36.4386    & -37.2318    & -37.3999    \\
Hexamer\cite{Bates2009}  & Prism       &   -45.92  & -46.4157    & -43.1065     &  -46.3284   \\
 & Cage                    &   -45.67  & -46.2443    & -44.0680     &  -42.7931   \\
 & Bag                     &   -44.3   & -45.0311    & -44.0609     &  -44.0417   \\
 & CyclicChair             &   -44.12  & -44.9882    & -46.4263     &  -45.8016   \\
 & Book A                  &   -45.2   & -45.8583    & -45.0931     &  -45.8630   \\
 & Book B                  &   -44.9   & -45.5591    & -44.4939     &  -44.7469   \\
 & CyclicBoat A            &   -43.13  & -43.9873    & -45.0067     &  -43.3573   \\
 & CyclicBoat B            &   -43.07  & -43.9275    & -44.8261     &  -43.8772   \\
Octamer\cite{Xantheas2004} & S$_4$        &   -72.7   & -73.3850    &  -72.5324    &  -70.1324   \\
 & D$_{2d}$                &   -72.7   & -73.4192     &  -72.5400    &  -69.6721   \\
11-mer\cite{Bulusu2006} & 434           &   -105.718& -103.54     & -104.2905    & -103.1903   \\
 & 515                     &   -105.182& -103.435    & -107.2288    & -104.4197   \\
 & 551                     &   -104.92 & -103.222    & -106.1787    & -104.2280   \\
 & 443                     &   -104.76 & -103.146    & -103.8261    & -104.0221   \\
 & 4412                    &   -103.971& -102.243    & -105.2387    & -105.0566   \\
16-mer\cite{Yoo2010} & Boat A        &   -170.8  & -164.220   &  -167.9238   &  -166.9812  \\
 & Boat B                  &   -170.63 &  -164.067  &  -167.2663   &  -166.7567  \\
 & Anti-boat               &   -170.54 &  -163.909  &  -166.4467   &  -166.7683  \\
 & ABAB                    &   -171.05 &  -163.969  &  -166.6278   &  -167.0233  \\
 & AABB                    &   -170.51 &  -163.413  &  -165.4890   &  -165.9624  \\
17-mer\cite{Yoo2010} & Sphere        &   -182.54 &  -175.323  &  -180.6810   &  -185.3541  \\
 & 5525                    &   -181.83 & -174.636   &  -181.1569   &  -183.1971  \\
20-mer\cite{Fanourgakis2004} & Dodecahedron  &   -200.1  &  -200.814  &  -204.1055   &  -206.4638   \\
 & FusedCubes              &   -212.1  &  -209.295  &  -208.6317   &  -209.4767  \\
 & FaceSharingPrisms       &   -215.2  &  -209.606  &  -212.1693   &  -211.8396  \\
 & EdgeSharingPrisms       &   -218.1  &  -211.476  &  -213.5750   &  -213.4915  \\
\hline
 &  &     MAD (Ref. Lit.) &  MAD (DFT)\cite{Mardirossian2014} & Units &    \\
\hline
MB-UCB-MDQ & Dimer to Octamer   & 0.759 &   0.788 & kcal/mol    &\\
 & 11-mer to 20-mer  &  2.770 & 2.865 & kcal/mol   &\\
 \hline
MB-UCB-MD & Dimer to Octamer   & 0.863 & 0.858  & kcal/mol  & \\
 & 11-mer to 20-mer  &  2.943 & 3.039& kcal/mol   &\\
 \hline
\end{tabular}}
\end{table}

\begin{figure}[H]
\includegraphics[scale=0.23]{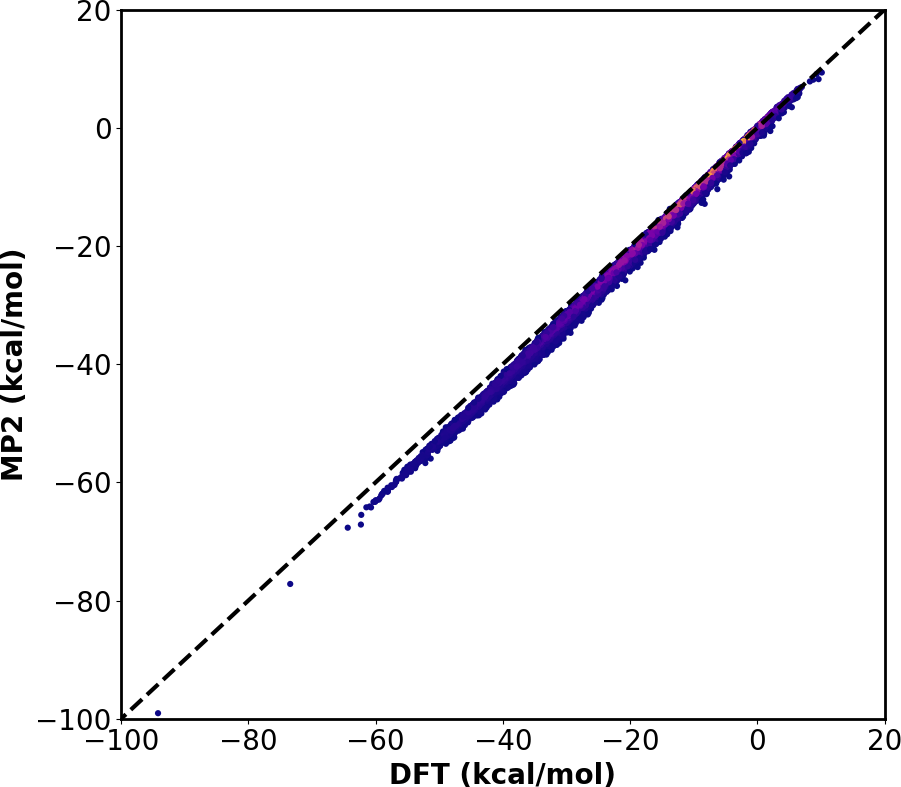}
\includegraphics[scale=0.23]{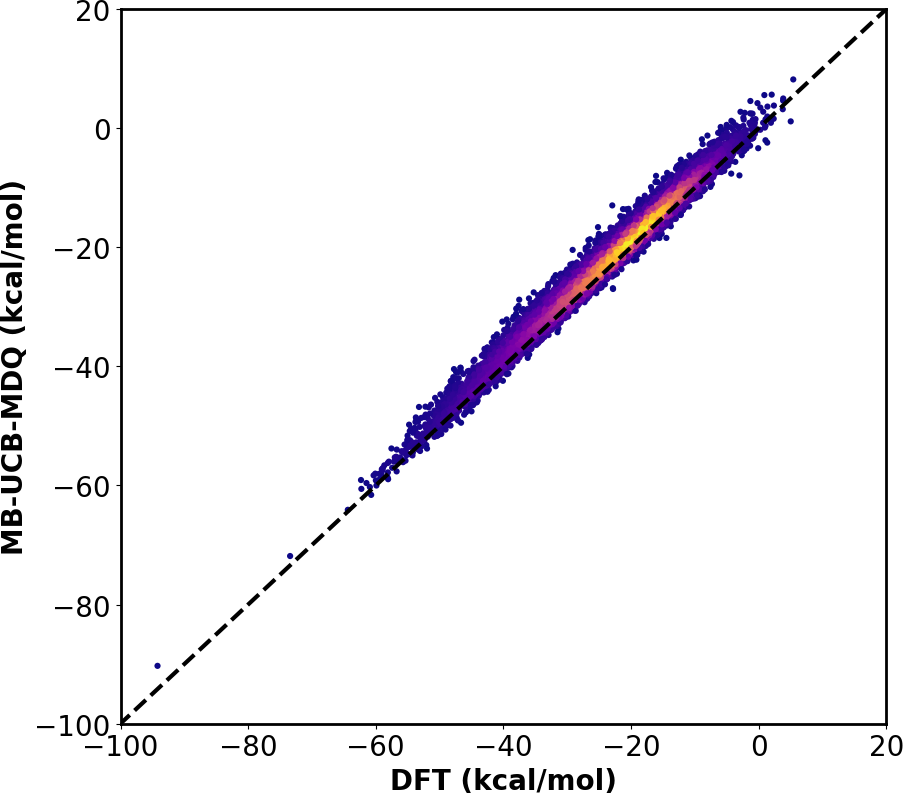}
\includegraphics[scale=0.23]{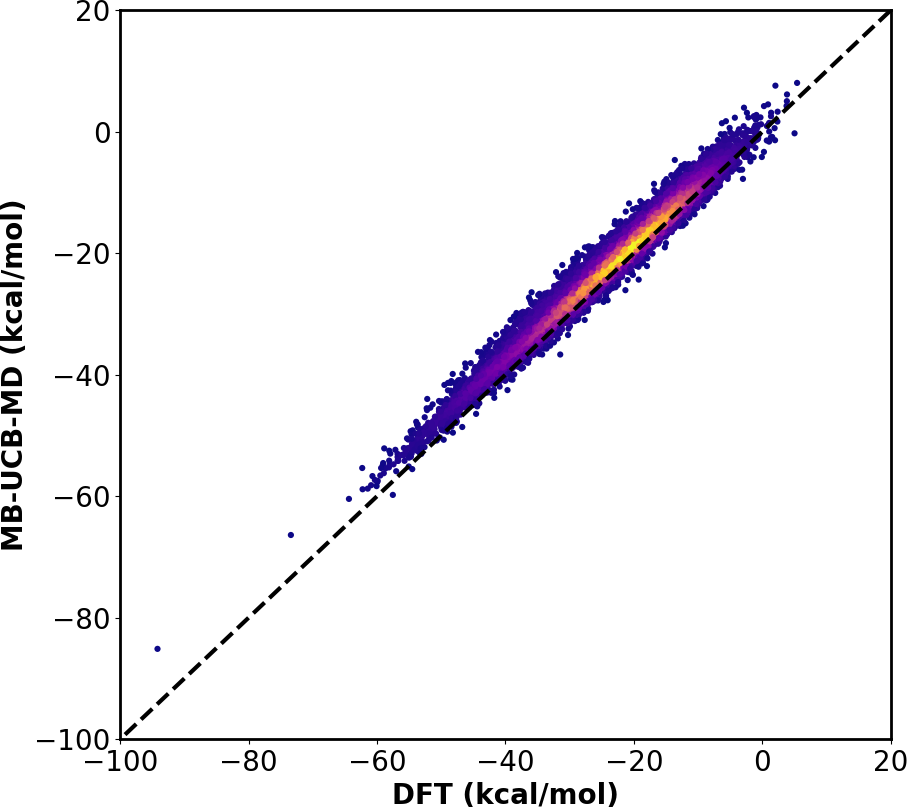}
\caption{Correlation of water cluster binding energies from dimer to decamers for various energy models. (a) comparison of water dimers to decamers for 21,604 configurations between RI-MP2/heavy-aug-cc-pVTZ and $\omega$B97X-V/def2-QZVPPD. Comparison from water hexamers to decamers for 12,002 configurations between $\omega$B97X-V/def2-QZVPPD and (b) MB-UCB-MDQ and (c) MB-UCB-MD. Density of points represented by blue (low) to yellow (high) in the heatmap.}
 \label{fig:fig5}
 \end{figure}

What is more relevant for the condensed phase is to consider distorted water cluster geometries taken from a liquid water simulation reported by Wang and co-workers\cite{Wang2013a,Laury2015}. Using 21,604 geometries for water clusters from dimers to decamers, we have developed a complete benchmarking suite using $\omega$B97X-V/def2-QZVPPD\cite{Mardirossian2014} which has itself been well-validated against a wide range of binding energies using the CCSD(T) gold standard\cite{Mardirossian2017}. When compared against the data set of the previously reported benchmarks based on RI-MP2/heavy-aug-cc-pVTZ, Figure \ref{fig:fig5}a indicates that the previous RI-MP2 results systematically overbind with respect to the DFT reference. The MB-UCB-MDQ (Figure \ref{fig:fig5}b) and MB-UCB-MD (Figure \ref{fig:fig5}c) models have a slight tendency to underbind with respect to the $\omega$B97X-V/def2-QZVPPD reference, although it is also evident that both perform equally well against the DFT benchmark. 

Of course the most important validation of any new water force field (and/or EDA scheme), is its characterization beyond the highly necessary tests on water cluster energies to its performance on condensed phase properties. Thus we have characterized the MB-UCB water model over a range of temperatures, confirming that it realizes a density maximum at 277 K, exhibits very accurate diffusion constants, and shows very good values for the heat of vaporization and heat capacity (in which both include quantum corrections\cite{Horn2004}) when compared to experiment (Figure 6a-6c and Table 2).

\begin{figure}[H]
\includegraphics[scale=0.08]{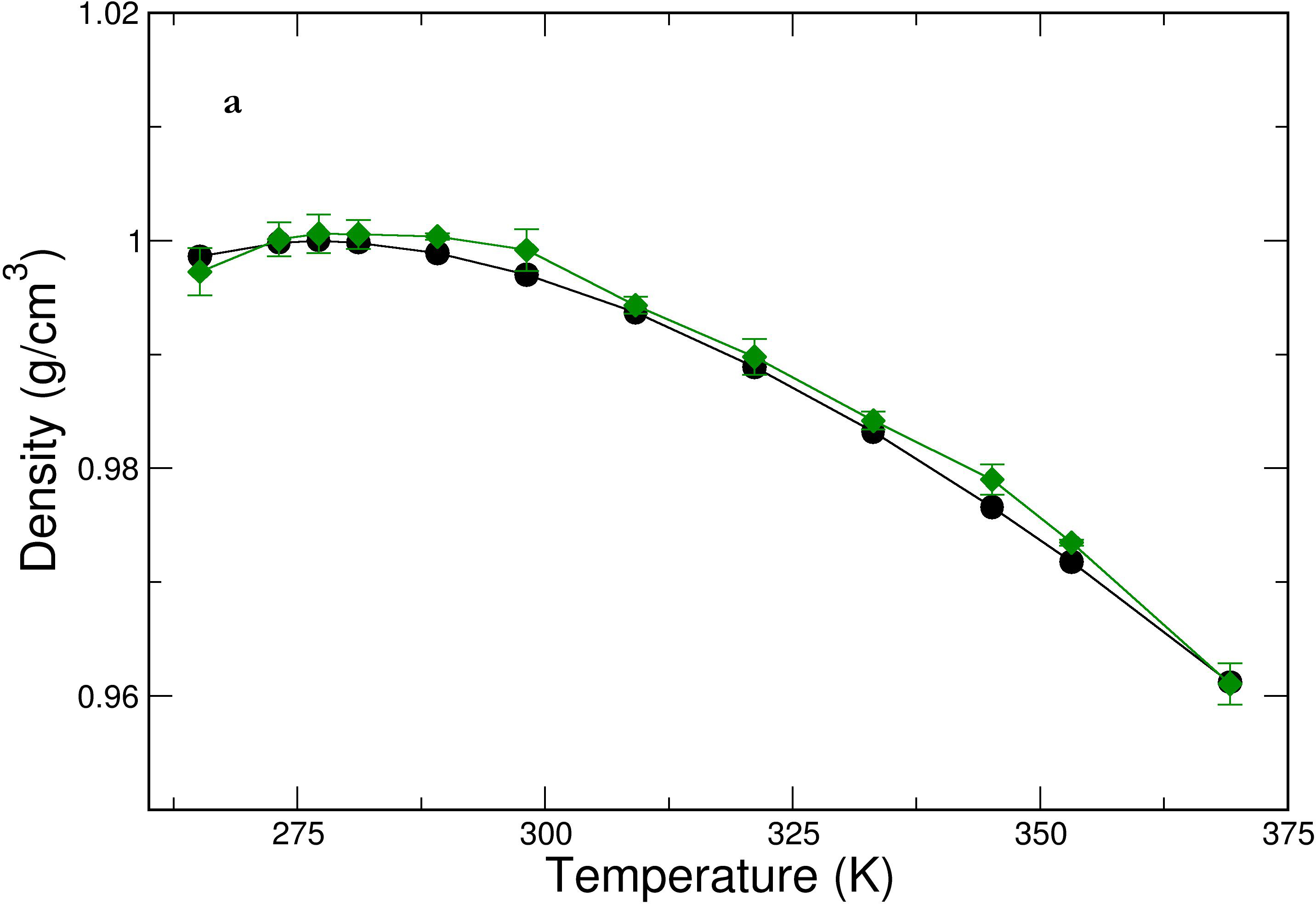}
\includegraphics[scale=0.08]{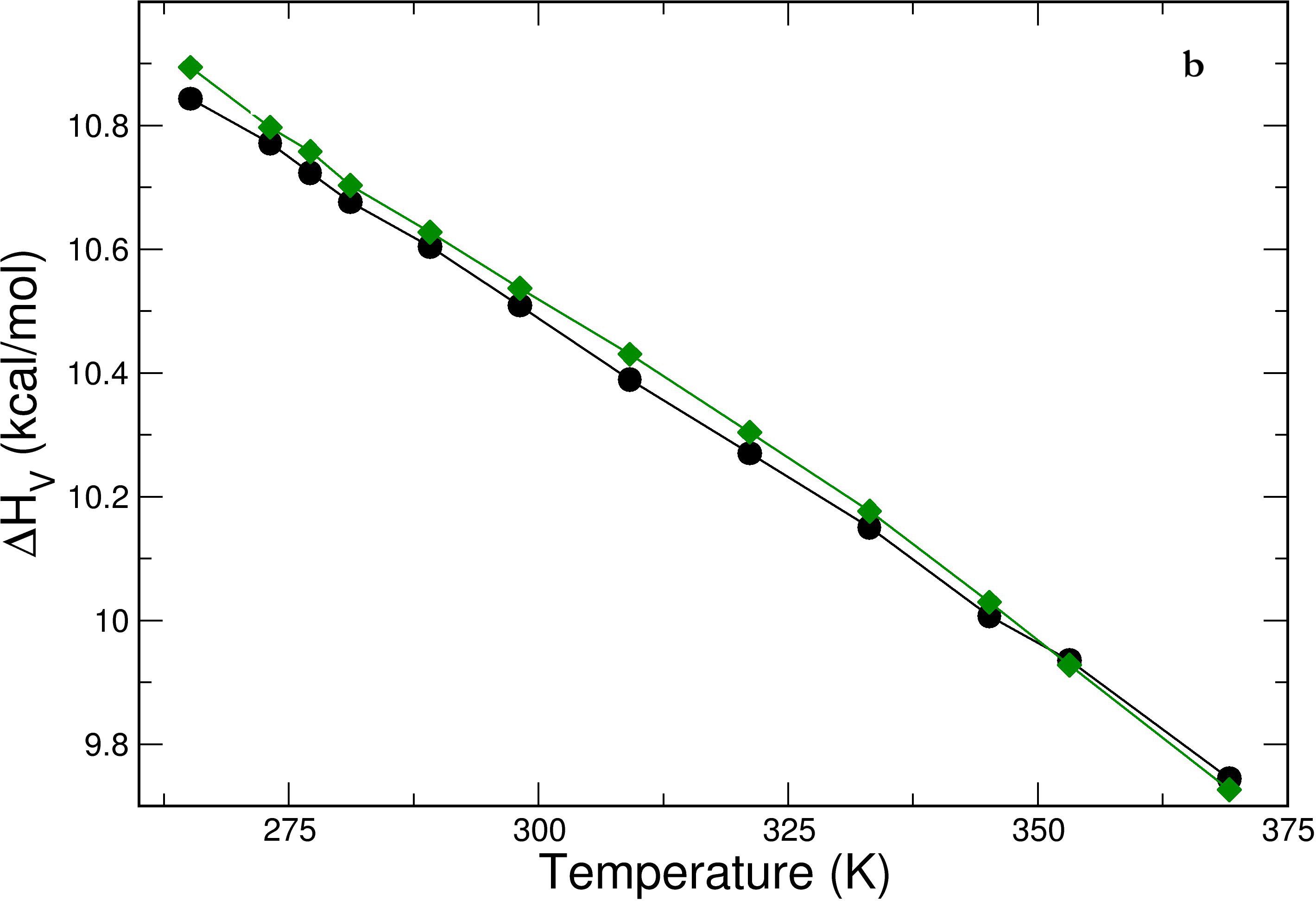}  
\includegraphics[scale=0.08]{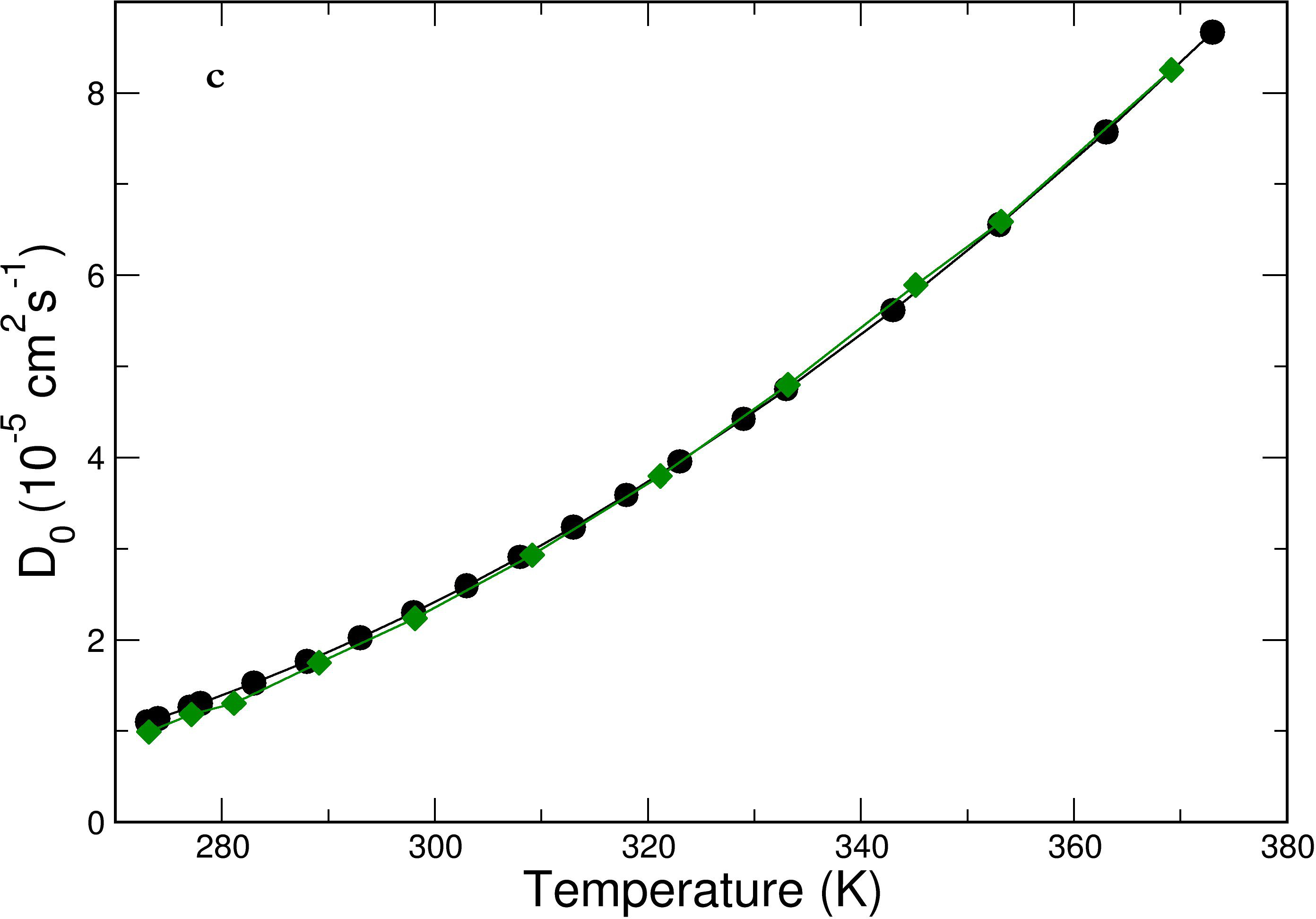}
\includegraphics[scale=0.08]{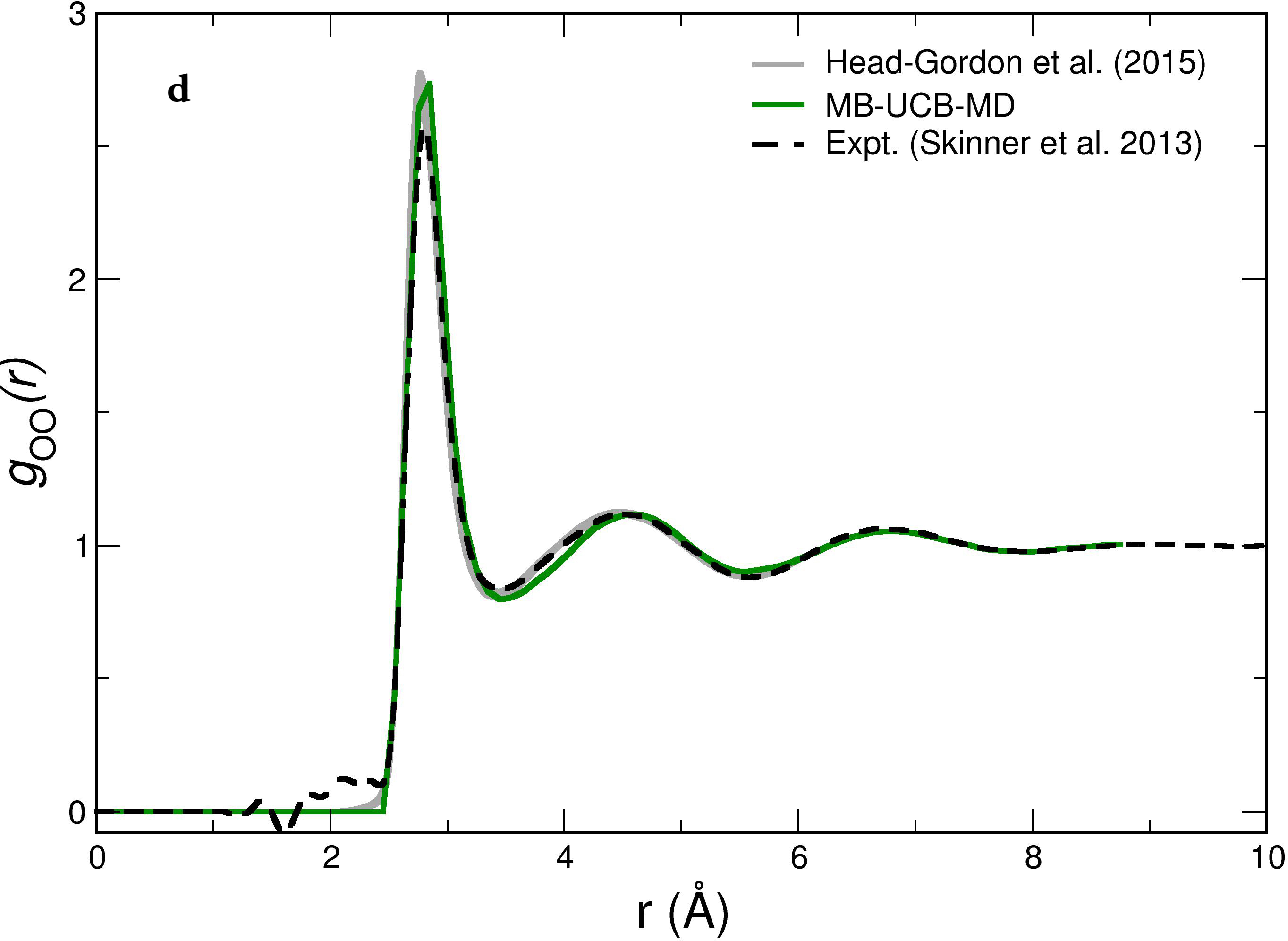}
\includegraphics[scale=0.08]{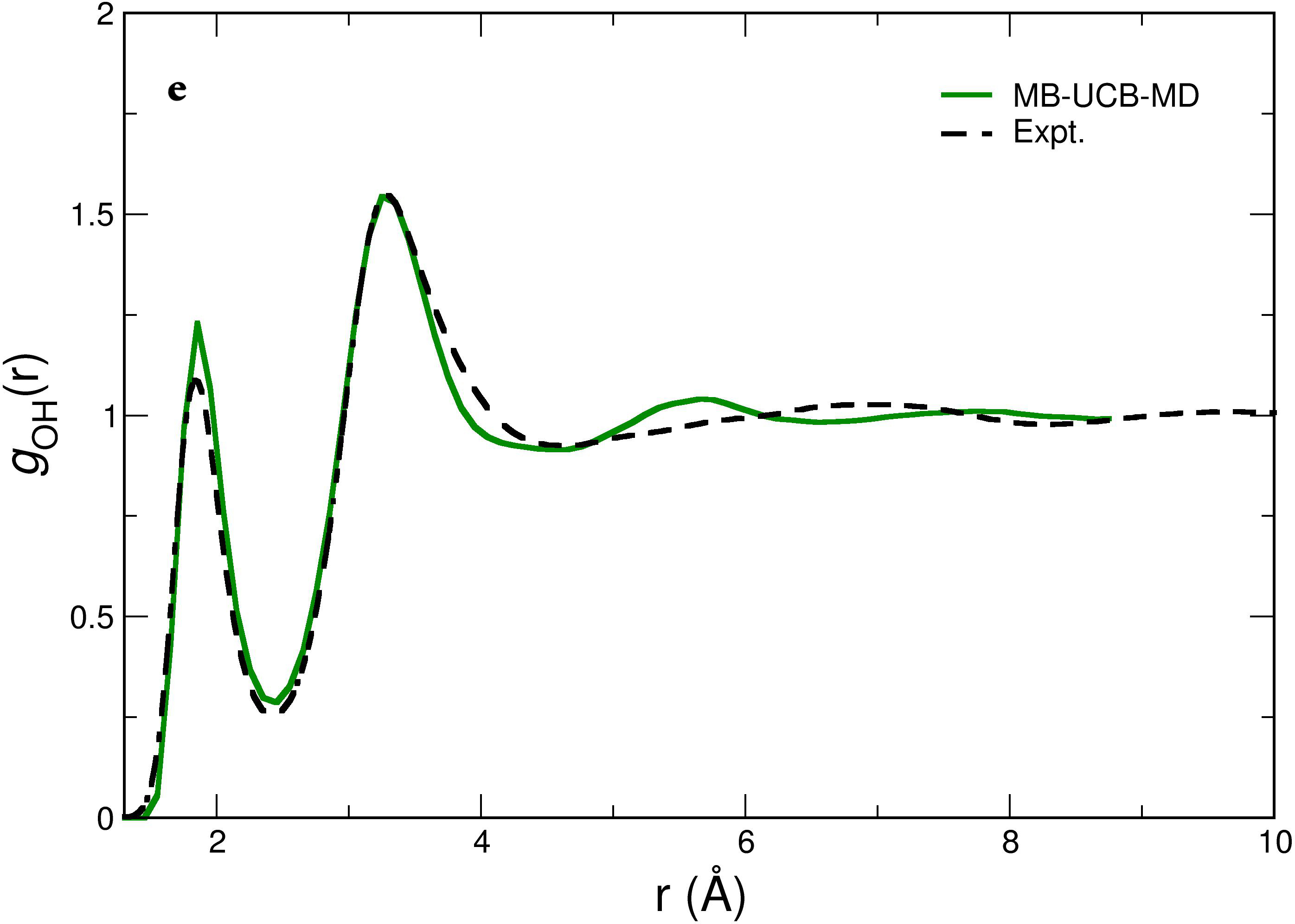}  
\includegraphics[scale=0.08]{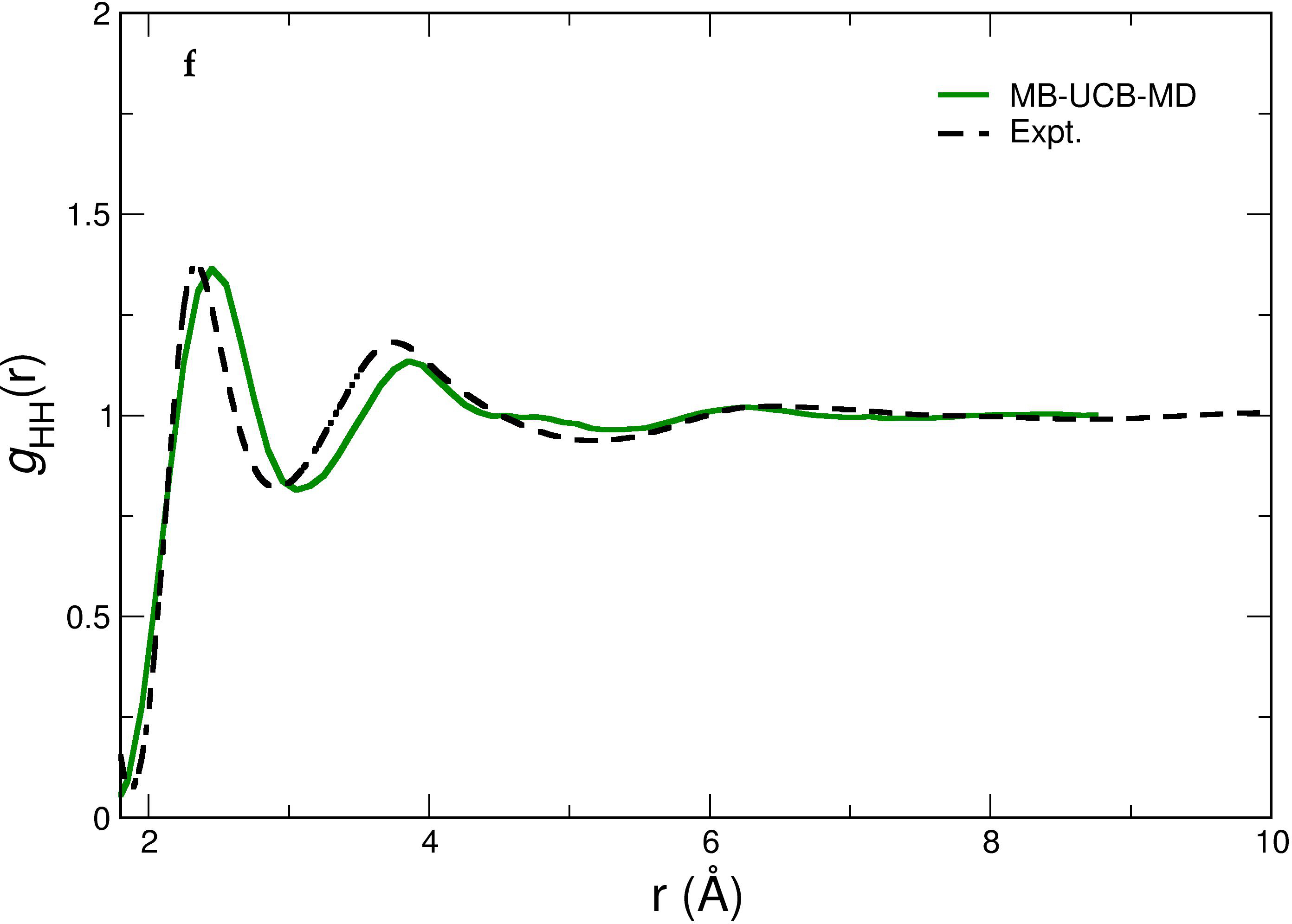}
\caption{Thermodynamic, transport, and structural properties for MB-UCB (green) vs. experiment (black). (a) density, (b) the diffusion coefficients which have been corrected for finite size effects using the experimental viscosity, and the (c) heat of vaporization. Tabulated data for density and heat of vaporization is given in Tables S1 and S2. (d-f) Radial distribution functions (rdfs) of water O--O, O--H,and H--H correlations. For gOO(r) the gray curves correspond to a family of allowed rdfs that remove the unphysical density at very low r and all conform to the isothermal compressibility\cite{Brookes2015}.}
\label{fig:properties}
\end{figure}

Given the correct density provided at room temperature, we have also calculated two additional properties at this one state point. The static dielectric constant of water was computed from the total dipole moment fluctuations of the system yielding a computed dielectric constant of 77(7) that agrees well with the experimental dielectric constant of 78.4\cite{Fernandez1997}. The MB-UCB-MD model is also in excellent agreement with the family of allowed gOO(r) functions\cite{Brookes2015} that are consistent with the experimental uncertainties inherent in the work of Skinner and co-workers\cite{Skinner2013}, and the gOH(r) and gHH(r) functions are also in notably good agreement with the neutron scattering studies of Soper\cite{Soper2000}. The MB-UCB-MD model, which is formulated on just small water cluster data from monomer to pentamers, offers significant improvement in liquid structure over previously reported iAMOEBA\cite{Wang2013a} and AMOEBA\cite{Ren2003,Laury2015} models that tend to be over-structured in spite of being fit to condensed phase data, something which has been less problematic for other water models such as MB-Pol\cite{Babin2012,Babin2013}, TIP4P-EW\cite{Horn2004}, and TIP5P\cite{tip5p}.

\begin{table}[H]
\caption{Isobaric heat capacity of liquid water at various temperatures obtained from the simulations using MB-UCB and compared with experiment. The heat capacity was derived by differentiating a 6th order polynomial fit to the heat of vaporization simulations, with error bars that are of the order of 1-2 cal$\cdot$mol$^{-1}$$\cdot$K$^{-1}$}.
\label{tab:heatcapcity}
\begin{tabular}{c|c|c|c}
\hline
Temperature (K) & MB-UCB & Quantum corrections\cite{Horn2004} & Experiment\cite{angell1982}\\
\hline
269.0 &  18.495 & -2.5 & 18.20\\
275.0 &  17.923 & -2.4 & 18.12\\
279.0 &  17.543 & -2.4 & 18.09\\
285.0 &  17.062 & -2.3 & 18.03\\
293.0 &  16.686 & -2.2 & 18.02\\
303.0 &  16.614 & -2.1 & 18.00\\
315.0 &  17.100 & -2.0 & 18.00\\
327.0 &  18.032 & -1.9 & 18.02\\
339.0 &  19.228 & -1.8 & 18.03\\
349.0 &  19.921 & -1.7 & 18.05\\
361.0 &  20.120 & -1.6 & 18.11\\
\hline
\end{tabular}
\end{table}

\section{Discussion and Conclusion}
The performance of all force fields rely on accurate representations of the individual energy components such as permanent electrostatics, many-body polarization, the van der Walls potential, and non-classical effects such as charge penetration and charge transfer to yield the total intermolecular energy. However many force fields  are parameterized with respect to either total energy and forces of high quality QM data and/or fits to experimental properties, and thus they are only indirectly connected to the individual energy components. In order to minimize errors among the individual energy terms of the decomposed QM energy, we have developed the MB-UCB force field based on the same breakdowns of  ALMO-EDA when analyzed on small water cluster data. By incorporating charge penetration\cite{Piquemal2003} to better describe the short range electrostatics, accompanied with atomic anisotropic polarizabilities to improve the polarization energy\cite{Das2018}, as well as introducing a simple model for the charge transfer energy,\cite{Shideng2017} we have found excellent agreement with the ALMO-EDA breakdown with respect to these energy components. Furthermore we have shown that we can eliminate the need for the algebra-intensive quadrupoles of the permanent electrostatics for water. 

As a result, we have shown that MB-UCB quantitatively describes the binding energies of both small and large water clusters compared to the {\it ab initio} binding energies. More importantly it has shown excellent properties outside the parameterization set including rdfs at room temperature, and excellent reproduction of temperature dependent thermodynamic and transport properties. We note that all condensed phase properties are true validation sets for the MB-UCB model, unlike the recent SAPT-guided AMOEBA+ model\cite{amoebaplus} whose parameters were optimized with ForceBalance across a large range of condensed phase properties. In fact, the MB-UCB model is much closer in philosophy to the MB-Pol model - which is highly accurate for water - by relying on small water cluster data and a quality hybrid DFT functional that is close to CCSD(T) accuracy, and our early examination of the temperature dependence of liquid state properties suggests that it has reached a similar level of accuracy. But we deviate from the MB-Pol model in several very important ways. First is that MB-UCB explicitly grapples with the quantum mechanical nature of the short-ranged two and three-body interactions, eliminating the need for the S2 and S3 terms that are heavily parameterized in the MB-Pol model (~1000 parameters for the three-body terms) and unique to each chemical species, thereby making MB-UCB far more tractable and transferable for chemical systems beyond water.

The limited existence of good functional forms to model charge transfer in force fields has been mainly due the fact that there has been no chemically motivated way of separating out the charge transfer energy in an electronic structure calculation. Therefore, most of the advanced force fields lack an explicit model for the charge transfer energy term or add it to polarization to reproduce SAPT induction\cite{amoebaplus}. However with the development of ALMO-EDA, which rigorously decouples polarization from charge transfer effects, it can provide guidance for incorporating charge transfer into an advanced force field. For the MB-UCB model we have used a highly empirical model for treating CT as an induction effect\cite{Shideng2017}, but augmenting it with a mutual induction term that introduces a greater many-body response. Although lacking the quantum mechanical effect of charge flow, the explicit representation of both CT and polarization is a step beyond the standard SAPT assumption of combining them into one induction term. In particular, because the exponential damping for CT is made more short-ranged than it is for polarization, we are able to correctly  describe the different spatial range dependencies of these two terms that captures an important aspect of their quantum mechanical differences. 

Because we have developed new algorithms to evaluate the many-body interaction energies and forces that reduces\cite{Albaugh2015} or eliminates\cite{Albaugh2017} the self-consistent field steps, the overall computational cost of the MB-UCB model is just that of the evaluation of the pairwise permanent electrostatics. Since the MB-UCB model only needs to utilize permanent atomic monopoles and dipoles for water, it should be computationally competitive with standard force fields but with greater accuracy afforded by its many-body character. The next stage of the MB-UCB force field development is for biologically relevant molecules starting with protein chemistry using amino acid building blocks, and ultimately extending the model to nucleic acids and drug molecules, and other non-biochemical systems as well.
 
\section*{Acknowledgment}
The authors thank the National Science Foundation for support under Grant No. CHE-1665315. This research used the computational resources of the National Energy Research Scientific Computing Center, a DOE Office of Science User Facility supported by the Office of Science of the U.S. Department of Energy under Contract No. DE-AC02-05CH11231. We would also like to thank Prof. Lee-Ping Wang for providing all the MD extracted water cluster geometries and corresponding MP2 water binding energies.
 
\bibliography{references}
\end{document}